\documentclass[english,aps,twocolumn,showpacs,amsmath,amssymb,prl,footinbib]{revtex4-2}
\usepackage[T1]{fontenc}
\usepackage[latin9]{inputenc}
\usepackage{amsmath}
\usepackage{graphicx}
\usepackage{esint}
\usepackage{lipsum}
\usepackage{bbold}

\makeatletter
\usepackage{graphicx}
\usepackage{color}
\usepackage{amsmath}
\usepackage{psfrag}
\usepackage[abs]{overpic}
\usepackage{enumitem}
\usepackage{natbib}
\usepackage{float}
\usepackage{verbatim}
\usepackage[normalem]{ulem}
\usepackage[%
  colorlinks=true,
  urlcolor=blue,
  linkcolor=blue,
  citecolor=blue
]{hyperref}

\makeatother

\usepackage{babel}

\begin{document}

\title{Fractional Josephson effect versus fractional charge in superconducting-normal
metal hybrid circuits}
\author{Mohammad Atif Javed$^{(*)}$, Jakob Schwibbert$^{(*)}$, and Roman-Pascal Riwar}
\affiliation{Peter Gr\"unberg Institute, Theoretical Nanoelectronics, Forschungszentrum
J\"ulich, D-52425 J\"ulich, Germany}

\begin{abstract}
Fractionally charged excitations play a central role in condensed matter physics, and can be probed in different ways. If transport occurs via dissipation-less supercurrents, they manifest as a fractional Josephson effect, whereas in dissipative transport they can be revealed by the transport statistics. However, in a regime where supercurrents and lossy currents coincide, a full understanding of the relationship between these two transport phenomena is still missing. Moreover, especially for superconducting circuits, the question of how noninteger quasicharges can be reconciled with charge quantization is still not fully resolved, and plays an important role for the circuit dynamics.
Here, we aim to unify the above concepts by studying the system-detector dynamics in terms of a Lindbladian capturing both coherent and dissipative transport. Charge quantization is here a conserved property of the detector basis of the Lindbladian, while charge fractionalization is a topological property of its complex-valued eigenspectrum.
We show that already conventional
superconductor-normal metal hybrid circuits exhibit
a variety of topological phases, including an open quantum system
version of a fractional Josephson effect. Surprisingly, 
quasiparticles, usually considered a detrimental side effect,
are here a necessary ingredient to observe nontrivial transport behaviour.  \\ \\ $(*)$ Both authors contributed equally to this work. 
\end{abstract}
\maketitle

\section{Introduction}

The notion of fractional charges plays an omnipresent role in condensed
matter physics, especially in lower dimensional systems, such as in
1D Luttinger liquids \cite{Pham_2000,Imura_2002,Trauzettel_2004,Steinberg_2007,Gutman2010},
or in the 2D fractional quantum Hall effect \cite{Tsui_1982,Laughlin_1983,Kane_1994,Saminadayar_1997,de-Picciotto_1997},
as well as in topological superconductors, where the presence of Majorana-
or parafermions gives rise to a fractional Josephson effect \cite{Kitaev_2001,Fu_2009,Zhang_2014,Orth_2015}.
While the literature on how to define and detect a fractional charge
$e^{*}\neq e$ ($e$ being the elementary charge) is of course much
vaster than we could possibly account for in this small introduction,
we can nonetheless identify two main and seemingly distinct flavours,
which we here intend to unify. As we argue below, this attempt of a unification is deeply rooted in the understanding, that fundamentally, charge of any electronic system must be quantized in integer units of the elementary charge $e$, such any charge fractionalization effect can only be meaningfully defined in terms of the topological properties of the time evolution of a system coupled to a transport detector.

In a non-equilibrium transport situation, $e^{*}$ may be extracted
from the transport statistics. This idea was pioneered by Kane and
Fisher \cite{Kane_1994}, who showed that the Fano factor (the noise-to-current
ratio) returns $e^{*}/e$, provided that the transport statistics
is Poissonian. Very recently, this idea was generalized to a
generic non-Poissonian transport regime, when considering the topological
properties~\cite{Riwar_2019b} of the entire full-counting statistics
(FCS)~\cite{Levitov_1996}. This definition hinges on the time-dependent dynamics of the moment
generating function $m\left(\chi\right)=\sum_{N}e^{i\chi N}P\left(N\right)$,
where $\chi$ is the so-called counting field, and $P\left(N\right)$
is the probability of having transported $N$ electrons. If $N$ is integer, then $m$ is obviously $2\pi$-periodic in $\chi$ for all times. Effective quantum field theories hosting fractionally charged excitations may in principle predict a moment generating function $m$ with broken periodicity. However, it was understood already, that the elementary charge being fundamental for any electronic system, this broken periodicity must be artificial. That is, such effective field theories must have a limited validity for sufficiently high cumulants, and the $2\pi$-periodicity of $m\left(\chi\right)$ must be restored
\cite{Aristov_1998,Gutman2010,Ivanov_2013,Riwar_2019b,Riwar_2021}. On the other hand,
generic open quantum systems with a transport detector were shown to undergo dynamical phase transitions
\cite{Ubbelohde_2012,Flindt_2013,Brandner_2017} leading to a braiding
of the complex eigenspectrum of the Lindbladian along the counting
field \cite{Ren_2012,Li_2013}. The main nontrivial contribution of~\cite{Riwar_2019b} is the realization, that the resulting breaking of the $2\pi$-periodicity
of the complex eigenspectrum, which governs the time evolution of $P\left(N\right)$, and thus of $m(\chi)$,
should be interpreted as transport carrying fractional charges in the same sense as the known examples from strongly correlated systems. In short, fundamental charge quantization is thus a property of the detector basis, whereas fractional charges are a property of the open system eigenspectrum. Based on this realization, Ref.~\cite{Riwar_2019b} argued that fractional charges are already observable for standard sequential electron tunneling through a quantum dot in a purely dissipative transport regime, not requiring any material-specific properties or interactions.

Fractional charges can also be defined without the explicit need for
nonequilibrium transport (and transport statistics measurements) by
the phase picked up when travelling through a magnetic field, $\phi=e^{*}\int dxA\left(x\right)/\hbar$,
where for anyonic excitations, $e^{*}$ can be directly linked to the nontrivial exchange statistics \cite{Leinaas_1977,Wilczek_1982,Arovas_1984,Bartolomei_2020,Nakamura_2020}.
This notion of fractional charges is at the heart of the fractional
Josephson effect due to the presence of exotic excitations, such as Majorana- or parafermions \cite{Kitaev_2001,Fu_2009,Zhang_2014,Orth_2015,Kapfer_2019}. Here, $e^{*}$ defines the periodicity with which the supercurrent
depends on the superconducting phase bias $\phi$, since in superconducting transport, the phase enters as $e^{\pm i\phi}$. For a pure superconducting regime, it is usually possible to describe the dissipation-free current in the form of a low-energy Hamiltonian $H(\phi)$. It may therefore be tempting to just equip $H$ with the periodicity in $\phi$ given by the fractional Josephson effect. However, also in the context of superconducting circuits, there is a lingering question about the importance of charge quantization in various contexts, such as for charge noise sensitivity of the fluxonium~\cite{Koch_2009,Manucharyan_2009,Mizel_2020}, when coupling Josephson junctions to an electromagnetic environment~\cite{Murani_2020,Hakonen_2021,Murani_2021,Kaur_2021}, or when considering the physics of quantum phase slip junctions~\cite{Koliofoti_2022}. Specifically for topological superconductors, the presence of Majorana fermions provides a degenerate ground state with even and odd fermion parity, allowing for coherent transport processes which transport a single elementary charge $e$ instead of the Cooper pair charge $2e$ in the ordinary Josephson effect. However, there is a fundamental incompatibility between charge stored in the topological part of the circuit (integer multiples of $e$) and the charge stored in the trivial parts of the device (integer multiples of $2e$). This interplay can lead to instabilities of the fractional Josephson effect for certain circuit configurations~\cite{vanHeck_2011}. The relevance of this incompatibility has also been recently studied for time-dependent driving and capacitive coupling~\cite{Kenawy_2022}, giving rise to a purely geometric correction term. Generally it was argued~\cite{Riwar_2021}, that for superconducting systems, the periodicity in $\phi$ of a Hamiltonian (describing a given circuit element) is defined by the unit of charge which a detector, a magnetic field or another circuit element couples to, whereas fractional Josephson effects, and the associated fractional charges, are defined in the periodicity of the eigenspectrum. This argument thus corresponds to a quantum mechanical counterpart to the statement made for purely dissipative systems in Ref.~\cite{Riwar_2019b}.

The question which has, to the best of our knowledge, not yet been
addressed, is how exactly the fractional Josephson effect and fractional charges measured in the transport statistics are related, and importantly,
how to generalize to a situation where dissipative non-equilibrium
currents and equilibrium supercurrents coexist. When combining supercurrents
and FCS \cite{Romito_2004}, the counting field $\chi$ appears as
a shift in $\phi$, suggesting that the fractional charge in the moment
generating function is directly inherited from the periodicity of
the Josephson relation. However, we show that the picture becomes
much more complex when including nonequilibrium currents, and that
in the most generic situation, charge fractionalization expresses
itself as exceptional points (EP) in the 2D space spanned by the independent
superconducting phase $\phi$ and the counting field $\chi$. Charge fractionalization in the current statistics and the fractional Josephson effect are thus in general related, but nonetheless distinct effects. Moreover, and similar in spirit to Ref.~\cite{Riwar_2019b}, we can show that in a generic open system context, no exotic materials are required to engineer topological phase transitions giving rise to fractional charges, and a fractional Josephson effect. Curiously, we find that poisoning due to out-of-equilibrium quasiparticles, usually a nuisance for superconducting circuits~\cite{Lutchyn_2005,Shaw_2008,Catelani_2011,Leppakangas_2012,Fu_2009,vanHeck_2011,Rainis_2012,Goldstein_2011,Budich_2012,Pekola2013}), is in the particular case studied here a necessary ingredient driving the topological transitions.

For concreteness, we consider a minimal heterostructure model of a
single-level quantum dot coupled to two phase-biased superconductors
(S) allowing for a supercurrent to flow, and an additional normal
metal (N), providing a nonequilibrium electron source. Quantum dot
heterostructures have been widely studied in the past both theoretically
\cite{Fazio_1998,Kang_1998,Clerk_2000,Cuevas_2001,Pala_2007,Sothmann_2010,Hiltscher_2011,Futterer_2013,Sothmann_2014,Weiss_2017}
and experimentally \cite{Herrmann_2010,Hofstetter_2010,Dirks_2011},
with a recently revived interest in connection with a possible probing
of the Higgs mode \cite{Heckschen_2022}. In particular, the inclusion
of a counting field has been discussed in Ref. \cite{Soller_2014}.

While all of our results have been obtained specifically for this
model, we believe that our findings regarding the connection between
the fractional Josephson effect and fractional charges are generic, so long as the dynamics is described by a Lindbladian.
In particular, we find that the aforementioned EPs give rise to phase
transitions which can carry a trivial or fractional charge in $\chi$
(along the lines of \cite{Riwar_2019b}) for different values of
$\phi$, and at the same time a conventional or fractional Josephson
effect for different values of $\chi$. Our work can thus be seamlessly embedded in a larger
currently ongoing effort to generalize the notion of topological phase
transitions to open quantum systems \cite{Rudner_2009,Rudner_2010,Diehl_2011,Bardyn_2013,Budich_2015,Rudner_2016,Engelhardt2017,Bardyn_2018,McGinley_2018,Edvardsson_2019,Kawabata_2019,McGinley_2019,Kastoryano_2019,Lieu_2020},
especially when expressed via EPs in the open system eigenspectrum
\cite{Heiss_2004,Heiss_2012,Kunst_2018,Kawabata_2019,Wu_2019,Mandal_2021,Bergholtz_2021},
by here assigning them the explicit role of generators of fractional
charges and a fractional Josephson effect.

Finally, we explicitly illuminate the role of transport detectors,
for different transport measurement schemes. For instance, a generic
model of a charge meter constantly entangling with the measured transport
processes \cite{Pluecker_2017prb,Pluecker_2017,Schaller_2009} suppresses
all supercurrents, but nonetheless provides a new type of fractional
transport phase, which was not yet predicted in Ref. \cite{Riwar_2019b}, consisting of a statistical mix of trivial and fractional
charges.
This is constrasted with a complementary understanding of FCSs, where
the cumulant generating function is reconstituted by measuring individual
cumulants of the current statistics (in the spirit of the FCS as defined
in Ref. \cite{Romito_2004}), where supercurrents persist, and the
aforementioned EP phase-transitions are (at least in principle) measurable.
However, because the materials in the here considered circuit are trivial, the fractional Josephson effect
is only visible at finite counting fields $\chi$, and its unambiguous
observation would thus in principle require the measurement of cumulants
of arbitrarily high order. In order to circumvent this issue, we study
alternatively quantum weak measurements of the supercurrent. While
weak measurement of the current could be envisaged by means of Faraday
rotation (explicitly proposed to weakly measure spins, e.g., in Ref.
\cite{Liu_2010}), we strive to propose an ``all-circuit'' realization
of weak measurement using SQUID detectors, inspired by Ref.~\cite{Steinbach_2001}.
We show in particular that a certain post-processing of the classical
information obtained by the weak measurement allows to simulate the
influence of a finite counting field, and thus induces the protected
fractional Josephson effect. This principle can to some extend be understood
as a new paradigm of the information of a weak detector being used
to ``filter out'' transport processes with integer Cooper pairs in
favour of fractional Cooper pair processes, importantly, without the
need of real-time feedback \cite{Vijay_2012}.

This work is organized as follows. In Sec.~\ref{sec:integer_versus_fractional_JE} we set the stage by reviewing generic features of conventional and fractional Josephson effects by comparing the quantum dot circuit with a Majorana-based circuit. In the same section we define a notion of an open system fractional Josephson effect in presence of a generic coupling to a bath. This is followed by Sec.~\ref{sec:normal_metal} where we explicitly introduce dissipation by means of a coupling between normal metal-induced dissipative mechanism for the quantum dot circuit. In Sec.~\ref{sec:different_flavors_of_FCS} two common versions of full-counting statistics are introduced and the relevant features with respect to topological transport are elucidated. In Sec.~\ref{sec:fractional_charge_vs_fractional_JE} we discuss various topological phase transitions which arise due to the interplay of dissipation and transport measurement, and argue how they can be interpreted as fractional charges and a fractional Josephson effect, respectively. Based on these results, we investigate in Sec.~\ref{sec:Weak_measurement} how continuous weak measurement of the current can serve as a means to reach the part of the phase space with finite counting field, where the fractional Josephson effect can be observed. The conclusions are presented in Sec.~\ref{sec_conclusions}. Finally, several appendices detail important derivations and intermediate results, such as the computation of the open system eigenmodes and their interpretation (Appendix~\ref{app:eigenmodes}), the calculation of the position of exceptional points (Appendix~\ref{app:excep_points}), the derivation of how to extract higher cumulants of all eigenmodes (Appendix ~\ref{app:analytic_continuation}), and the calculation of the scattering properties of the SQUID detector used for weak measurement (Appendix~\ref{app:transmission_coeff}).

\section{Integer versus fractional Josephson effect and open quantum system
generalization\label{sec:integer_versus_fractional_JE}}

\begin{figure}[htbp]
\centering\includegraphics[width=1\columnwidth]{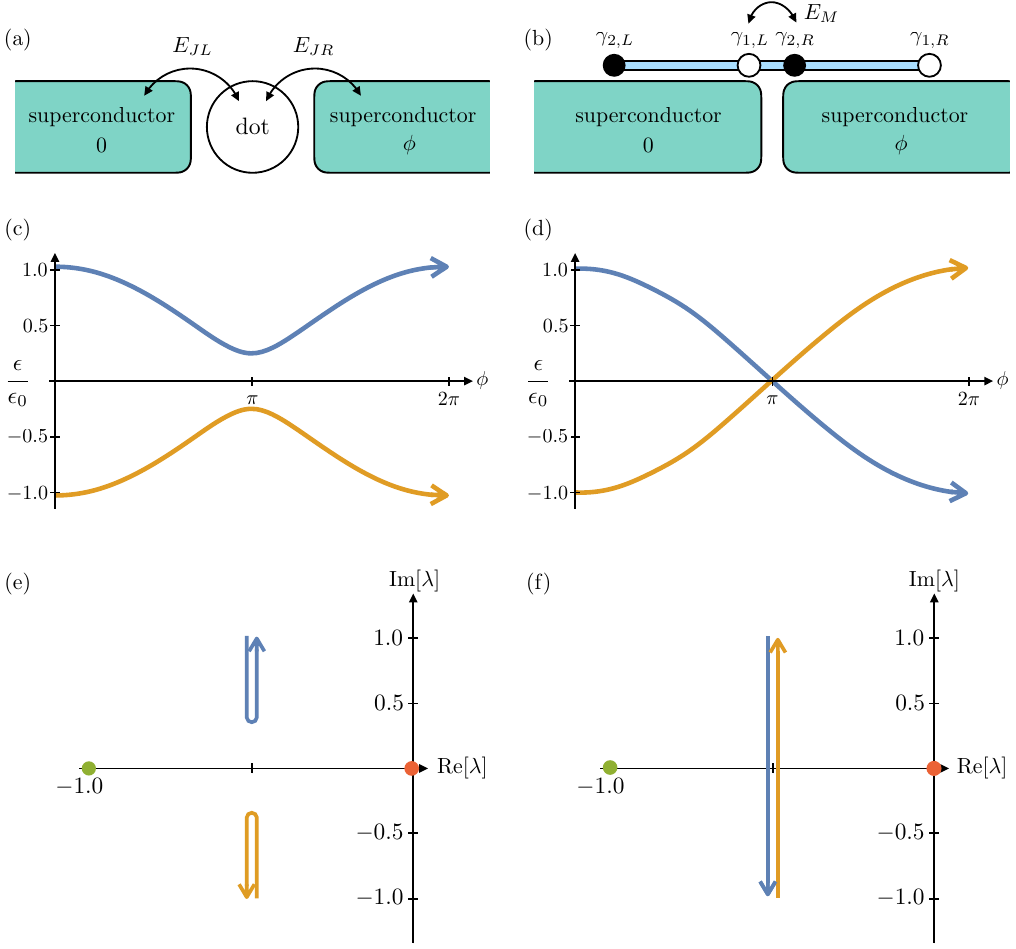}

\caption{Generalization of integer versus fractional Josephson effect for open
quantum systems. We compare a conventional circuit, such as the quantum dot proximitized with two superconductors (essentially a Cooper pair transistor) (a), with a topological circuit, hosting Majorana
fermions (b). Both circuits are subject to the phase bias $\phi$.
For the closed system, the regular and fractional Josephson effect can be distinguished by either a $2\pi$-periodic (c), or a $4\pi$-periodic (d)  energy
spectrum as a function of $\phi$.
When
including a simple, generic open system dynamics, see Eq.~\eqref{eq_Lindblad_generic}, the purely real spectrum in (c,d)
gets replaced by a complex spectrum $\lambda\left(\phi\right)$ (e,f),
where the real part describes dissipation and decoherence, whereas
the imaginary part represents the coherent dynamics. When drawing the complex spectrum of
the open system parametrically with respect to $\phi$ (from $0$
to $2\pi$, in the sense indicated by the arrow), we see that in the regular Josephson
effect, the eigenvalues with finite imaginary part return to their initial values
after a progression of $\phi$ from 0 to $2\pi$ (e). For the fractional
Josephson effect, these two eigenvalues swap places (f). \label{fig:open_FJE}}

\end{figure}

Consider two superconducting contacts with a phase difference $\phi$ (which may be controlled, e.g., by a magnetic field).
These superconductors may be brought into electrical contact through
various ways, for instance by an insulating barrier (the SIS
junction)~\cite{Josephson_1974} or via more general weak links~\cite{Beenakker_1991}. In this work, we consider for concreteness weakly tunnel coupled junctions,
which have a single quantum dot level~\cite{Konig_thesis,foot_1} sandwiched in between, see Fig.~\ref{fig:open_FJE}a. The quantum dot itself is described by a single level at energy
$\epsilon$, which can be at most doubly occupied (due to spin degeneracy).
The double occupation comes in addition with the energy cost $U$
due to Coulomb interactions. The Hamiltonian thus reads
\begin{equation}
H_{\text{dot}}=\epsilon\widehat{n}+U\frac{\widehat{n}\left(\widehat{n}-1\right)}{2},
\label{eq:quantum_dot_hamiltonian}
\end{equation}
the total occupation number on the single level being $\widehat{n}=\sum_{\sigma}d_{\sigma}^{\dagger}d_{\sigma}$,
where $d_{\sigma}^{\left(\dagger\right)}$ annihilates (creates) an
electron with spin $\sigma=\uparrow,\downarrow$. The eigenstates
of $H_{\text{dot}}$ are $\left|0\right\rangle $, $\left|1\right\rangle _{\sigma}=d_{\sigma}^{\dagger}\left|0\right\rangle $,
and $\left|2\right\rangle =d_{\uparrow}^{\dagger}d_{\downarrow}^{\dagger}\left|0\right\rangle $,
where $\left|0\right\rangle $ is the empty state, $d_{\sigma}\left|0\right\rangle =0$.
The coupling to the superconductors will in leading order introduce
coherent transitions between the $\left|0\right\rangle $ and $\left|2\right\rangle $
states, such that the transport can be captured in terms of the Hamiltonian
\begin{equation}
H\left(\phi\right)=H_{\text{dot}}+H_{J}\left(\phi\right),
\label{eq:complete_hamiltonian}
\end{equation}
where the exchange of Cooper pairs is described by 
\begin{equation}
H_{J}\left(\phi\right)=\frac{E_{JL}+E_{JR}e^{i\phi}}{2}d_{\uparrow}^{\dagger}d_{\downarrow}^{\dagger}+\text{h.c.}\label{eq:Josephson_potential_energy}
\end{equation}
The origin of this additional pairing term is the proximity effect, here considered in the limit of large superconducting
gaps $\Delta$ \cite{Rozhkov_2000,Vecino_2003,Choi_2004,Futterer_2009,Sothmann_2010}, such
that $E_{J\alpha}=\Gamma_{S\alpha}$ where $\Gamma_{S\alpha}$ is
the normal state tunneling rate between the quantum dot and the corresponding
contact $\alpha=\text{L,R}$. That is, the relevant correlation time of the superconducting reservoir is $\Delta^{-1}$.
In some sense, the tunnel coupling to the superconductor
reservoirs already represent an opening of the local quantum system to a reservoir. However, since supercurrents are mediated entirely without dissipation (at least in this approximation) this effect can be captured by
a low-energy Hamiltonian. Hence, the dissipation-free circuit constitutes our ``closed'' quantum system.

While these individual processes each give rise
to single Cooper pair tunneling processes ($\sim e^{\pm i\phi}$),
the presence of the quantum dot level modifies the overall transport
behaviour. Namely, the Hamiltonian $H\left(\phi\right)$ has the even
eigenstates $\left|\pm\right\rangle =\frac{1}{\sqrt{2}}\sqrt{1\pm\delta}\left|0\right\rangle \pm\frac{1}{\sqrt{2}}e^{i\phi_{J}}\sqrt{1\mp\delta}\left|2\right\rangle $
with
\begin{equation}
\delta=-\frac{2\epsilon+U}{\sqrt{\left(2\epsilon+U\right)^{2}+\left|E_{JL}+E_{JR}e^{i\phi}\right|^{2}}},
\label{eq:detuning}
\end{equation}
and the corresponding eigenenergies
\begin{equation}
\epsilon_{\pm}\left(\phi\right)=\frac{-2\epsilon-U\pm\sqrt{\left(2\epsilon+U\right)^{2}+\left|E_{JL}+E_{JR}e^{i\phi}\right|^{2}}}{2}.\label{eq:spectrum_quantum_dot}
\end{equation}
Since the odd parity states $\left|1_{\sigma}\right\rangle $ cannot partake
in the Cooper pair transport, they remain eigenstates also for the
full $H$, with the unchanged eigenenergy $\epsilon$. The above energies $\epsilon_{\pm}\left(\phi\right)$
are no longer a pure cosine (as for instance for the standard Josephson effect), but the spectrum remains in general $2\pi$-periodic, see Fig. \ref{fig:open_FJE}c, signifying integer multiple Cooper pair tunnelings
as explained initially in this paragraph.

There is however one special point in parameter space, where the $2\pi$-periodicity
is broken: for $\epsilon=-U/2$ and $E_{JL}=E_{JR}\equiv E_{J}$,
the minigap closes, such that $\epsilon_{\pm}=\pm E_{J}\cos\left(\phi/2\right)$, and the eigenvalues exchange places when progressing $\phi$ by $2\pi$, see Fig.~\ref{fig:open_FJE}d.
Here, it seems that the transport can be described by means of a fractional transport
of Cooper pairs, transferred in half-integer portions. 

This is highly reminiscent of topological
Josephson junctions based on Majorana fermions, where the fractional
Josephson effect hinges upon the topological contacts having an even and odd ground state~\cite{Fu_2009}. Here,
the Hamiltonian is commonly given in the form
\begin{equation}\label{eq:majorana_H}
    H_{M}=iE_{M}\cos\left(\phi/2\right)\gamma_{1,\text{L}}\gamma_{2,\text{R}}\ ,
\end{equation}
describing the coupling of Majorana edge states on the left and right
$\gamma_{1/2,\alpha}$ via a junction, see Fig. \ref{fig:open_FJE}b (see, e.g., Ref.~\cite{Leijnse_2012}). This Hamiltonian has the exact same $4\pi$-periodic eigenvalues $\epsilon_\pm(\phi)=\pm E_M \cos(\phi/2)$ (see again Fig.~\ref{fig:open_FJE}d), which we associate to the eigenvectors $|\pm\rangle$~\footnote{We note that for the open Majorana circuit, each of the eigenvalues is two-fold degenerate, corresponding to overall even and odd parity. For our purposes, this distinction is of no further importance.}. Note that of course, the eigenstates $|\pm\rangle$ of the Majorana circuit are different from the eigenstates $|\pm\rangle$ of the quantum dot circuits. We nonetheless choose the same notation for simplicity - the reason for this will become obvious below.

Now, the reader might perhaps be surprised by such seemingly naive (or even slightly brazen) juxtaposition of a regular quantum dot circuit and a Majorana-based junction. Indeed, one might for instance argue that no experiment could ever tune both $E_{JL,JR}$ and $\epsilon$ to such perfection as to make the mini-gap disappear completely. However, it should be noted that in Majorana circuits, finite size effects are known to induce a small gapping, due to a coupling of the Majoranas on the same chain (i.e., terms of the form $\sim \gamma_{2,\text{L}}\gamma_{1,\text{L}}$ or $\sim \gamma_{2,\text{R}}\gamma_{1,\text{R}}$)~\cite{Kitaev_2001,Lutchyn_2010,Wang_2022}. A gapping and thus a restoring of a $2\pi$-periodic spectrum was also predicted when a Majorana-junction and a regular Josephson junction are coupled in parallel to form a SQUID~\cite{vanHeck_2011}. As a consequence, the line between fractional and regular Josephson effect starts to blur, as a minigap may likely be present for both trivial and topological circuits.

Alternatively, one might have tried to argue that while the energy spectrum of both systems looks similar, the Hamiltonian has a fundamentally different periodicity in $\phi$, with Eq.~\eqref{eq:Josephson_potential_energy} being $2\pi$-periodic whereas Eq.~\eqref{eq:majorana_H} appears genuinely $4\pi$-periodic. Such arguments can however likewise be easily defused. If the phase bias $\phi$ is stationary, the periodicity of the Hamiltonian is a simple gauge choice and not of relevance. For instance, we could have redistributed the phase drop in Eq.~\eqref{eq:Josephson_potential_energy} symmetrically over both junctions with a factor $e^{\pm i\phi/2}$, thus achieving a $4\pi$-periodic Hamiltonian. However, this basis choice becomes relevant (i.e., it ceases to be a mere gauge choice) if $\phi$ becomes time-dependent due to driving with magnetic fields~\cite{You2019,Riwar_2022}, or a dynamical quantum operator due to the addition of a capacitor~\cite{Kenawy_2022}, or if non-local correlation measurements are performed~\cite{Riwar_2021}. Here, the appropriate rule of thumb~\cite{Riwar_2021} is that the relevant charge (and the corresponding charge unit) is the charge that a magnetic field, a capacitor, or a detector couple to. For instance, for the Majorana circuit, Eq.~\eqref{eq:majorana_H}, a detector could measure the charge being transported across the actual Majorana junction, marked with an arrow in Fig.~\ref{fig:open_FJE}b. Then, the $4\pi$-periodic basis choice given in Eq.~\eqref{eq:majorana_H} is correct, since the Majorana wires physically exchange the charge $e$. If the same detector would however measure the charge entering one of the s-wave superconductor bulks, which proximitize the topological nanowire (the green bulks in Fig.~\ref{fig:open_FJE}b), then the correct basis choice must be a $2\pi$-periodic one (e.g. via the unitary transformation proposed by Ref.~\cite{vanHeck_2011}), to account for the fact, that the trivial, s-wave part of the circuit can in its ground state only accept integer Cooper pairs with charge $2e$.

The above short review leads us to two conclusions. First, whether a system provides a fractional or regular Josephson effect should be best described exclusively by the periodicity of the eigenspectrum, and not of the Hamiltonian, as the latter is either a gauge choice (for constant phase bias), or fixed by external factors. Second, even if a system consists of topological superconductors, there are many nontrivial factors that may lead to an instability of the fractional Josephson effect, by introducing a minigap, and restoring $2\pi$-periodicity. In the following, we will show with the example of the trivial quantum dot system, that the inclusion of a non-equilibrium quasiparticle reservoir, and the addition of a transport detector can as a matter of fact undo the gapping, and restore an open system version of the fractional Josephson effect, without the need for fine-tuning any system parameters. As already stated in the introduction, this effort will furthermore shed light on the intricate relationship between the fractional Josephson effect and fractional charges, as defined in the transport statistics~\cite{Riwar_2019b}.

However, before continuing, we need to develop as a next preparatory step a generalization of the notion of a fractional Josephson effect for open systems. To this end, we consider either the quantum dot circuit, Eq.~\eqref{eq:complete_hamiltonian}, or the Majorana circuit described in Eq.~\eqref{eq:majorana_H}, and add a generic open system dynamics to it. For this purpose, we take as the basis the two closed system
eigenstates $\left|\pm\right\rangle $ with eigenenergies $\epsilon_{\pm}\left(\phi\right)$,
where $\epsilon_{\pm}$ can now either be $2\pi$-periodic or $4\pi$-periodic, see Fig. \ref{fig:open_FJE}c and d. In the next section,
we discuss a concrete model for open system dynamics for the quantum dot circuit; here,
on this general, illustrative level, we are merely concerned with
simple generic open system processes, which consists of stochastic
transitions between the $\left|+\right\rangle $ and $\left|-\right\rangle $
states. Such processes can be described by a Lindblad quantum master
equation for the density matrix $\rho=\sum_{\eta,\eta'=\pm}P_{\eta \eta'}\left|\eta\right\rangle \left\langle \eta'\right|$,
\begin{equation}\label{eq_Lindblad_generic}
\dot{\rho}=-i\left[H\left(\phi\right),\rho\right]+\sum_{j=\pm}\Gamma_{j}\left(a_{j}\rho a_{j}^{\dagger}-\frac{1}{2}\left\{ a_{j}^{\dagger}a_{j},\rho\right\} \right),
\end{equation}
with $a_{\pm}=\left|\pm\right\rangle \left\langle \mp\right|$, such
that $\Gamma_{\pm}$ represents the rate for a stochastic jump from
$\left|\mp\right\rangle $ to $\left|\pm\right\rangle $. 

While in the closed system, we had the real eigenvalues $\epsilon_{\pm}\left(\phi\right)$,
the Lindblad equation gives rise to a set of complex eigenvalues $\left\{ \lambda_{n}\right\} $,
describing both the coherent and dissipative dynamics, see Fig. \ref{fig:open_FJE}e
and f. In addition, while the closed system could be described by
just two eigenvalues, due to the two degrees of freedom $+$ and $-$,
for the open system, we have four eigenvalues due to the enlarged
structure of the density matrix. The eigenvalue $0$ represents the fact that there is a unique
stationary state, $\rho_{\text{st}}=\left(\Gamma_{-}\left|-\right\rangle \left\langle -\right|+\Gamma_{+}\left|+\right\rangle \left\langle +\right|\right)/\left(\Gamma_{+}+\Gamma_{-}\right)$.
The eigenvalue $-\left(\Gamma_{+}+\Gamma_{-}\right)$ corresponds
to the decay of the diagonal density matrix elements (if the states $\left|\pm\right\rangle $
encode a qubit, this would be the $T_{1}$ time). Finally, there
are the eigenvalues $\pm i\left(\epsilon_{+}-\epsilon_{-}\right)-\left(\Gamma_{+}+\Gamma_{-}\right)/2$,
belonging to the eigenoperators $\left|\pm\right\rangle \left\langle \mp\right|$
which describe the coherent oscillations, including the decoherence
rate $\left(\Gamma_{+}+\Gamma_{-}\right)/2$ ($T_{2}$ time). Only this last pair of eigenvalues depends on $\phi$, such that the integer and fractional Josephson effects
can be characterized by means of their $\phi$-dependence. In order to represent the (now) complex eigenspectrum, we choose a parametric plot, where the real and imaginary parts of $\lambda$ are shown as two independent axes, and the resulting curves are parametrized by $\phi$ (in Figs.~\ref{fig:open_FJE}e and f). In the regular Josephson effect the
eigenvalues with finite imaginary part (coherent dynamics) map onto
themselves when running $\phi$ from $0$ to $2\pi$, see Fig.
\ref{fig:open_FJE}e. This is in contrast to the fractional Josephson
effect, where the same two eigenvalues swap places, see Fig. \ref{fig:open_FJE}f. The two resulting open system spectra are thus still topologically distinct, as one cannot continuously map from one to the other. This allows for a straightforward topological classification of the open system dynamics.

\section{The model: superconductor-normal metal hybrid circuit\label{sec:normal_metal}}

\begin{figure}[htbp]
\centering\includegraphics[width=0.7\columnwidth]{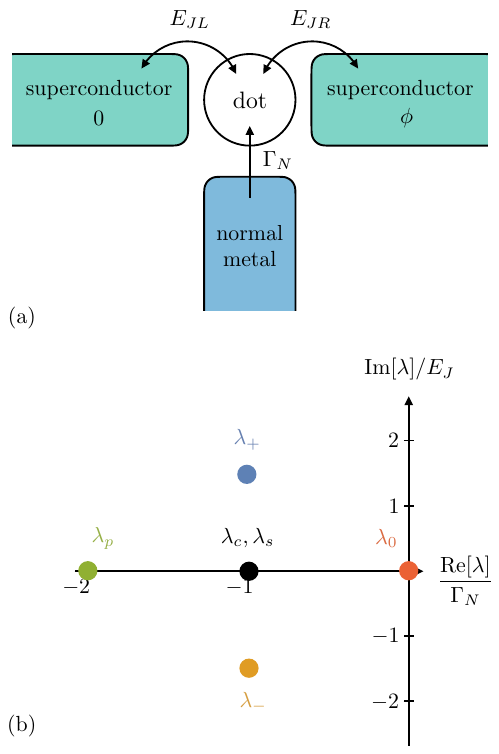}

\caption{(a) Sketch of the system under consideration. A central charge
island (quantum dot) is connected to a left and right superconductor,
with a phase bias $\phi$. The normal metal pumps quasiparticles into
the system with the rate $\Gamma_{N}$. A detector with counting field
$\chi$ measures current into the right superconductor. (b) The complex
eigenspectrum of the quantum master equation $\left\{ \lambda\right\} $
for $\chi=0$. The eigenmodes can be interpreted as follows. There
is a stationary state related to the eigenvalue $\lambda_{0}=0$.
The nonzero eigenvalues can be associated to the decay of the parity,
pseudocharge (see main text) and spin, $\lambda_{p,\overline{c},s}$,
respectively, and to the coherent dynamics $\lambda_{\pm}$.\label{fig:system}}

\end{figure}

Let us now introduce a microscopic model for the open system dynamics of the quantum dot circuit. In addition to the two superconductors, we include a tunnel coupling to a third, normal metal reservoir, see Fig. \ref{fig:system}a.
This coupling introduces dissipative, stochastic transport events.
Since pairing is absent in the normal metal, it can in lowest order
only introduce processes which flip the parity within the island.
We here focus on the regime where the chemical potential of the normal
metal is large with respect to the system dynamics ($\mu\gg\epsilon,U,E_{J\alpha}$).
Therefore, for the computation of the dynamics due to the normal metal
(by means of a standard sequential tunneling approximation, see Ref.~\cite{Konig_thesis}),
we may disregard the internal coherent dynamics. Consequently, the
normal metal will mainly act as a source of quasiparticles, thus inducing
the nonequilibrium stochastic transitions, $\left|0\right\rangle \rightarrow\left|1_{\sigma}\right\rangle $
and $\left|1_{\sigma}\right\rangle \rightarrow\left|2\right\rangle $.
Including the stochastic processes, the dynamics of the system is
described by the quantum master equation $\dot{\rho}=L\rho$, with
\begin{equation}
L\left(\phi\right)\cdot=-i\left[H\left(\phi\right),\cdot\right]+W_{N}\cdot,
\label{eq:kernel_qu_sc_metal}
\end{equation}
where $H(\phi)$ is the Hamiltonian given in Eq.~\eqref{eq:complete_hamiltonian}, and the kernel due to the normal metal processes, $W_N$, is of the
Lindblad form
\begin{equation}
W_{N}\cdot=\Gamma_{N}\sum_{\sigma}\left(d_{\sigma}^{\dagger}\cdot d_{\sigma}-\frac{1}{2}\left\{ d_{\sigma}d_{\sigma}^{\dagger},\cdot\right\} \right).
\end{equation}
We note that strictly speaking, the superconductors themselves should likewise contribute
to parity switches due to a finite quasiparticle population. However,
even though quasiparticles are known to occur with a much higher concentration
that what is expected from a thermal equilibrium distribution \cite{Martinis2009,Riste2013},
they are nonetheless dilute (with a concentration of typically $10^{-6}\sim10^{-5}$
with respect to the Cooper pair density \cite{Riwar2019}), such
that the normal metal influence may be expected to be dominant, even
when $E_{J}>\Gamma_{N}$, especially due to the large chemical
potential.

The processes $\left|0\right\rangle \rightarrow\left|1_{\sigma}\right\rangle $
and $\left|1_{\sigma}\right\rangle \rightarrow\left|2\right\rangle $
 both occur with the same rate. This is in particular due to $\mu\gg U$,
such that effectively, the many-body interaction is no longer visible
within the dissipative dynamics. This is why, for the remainder of this work, we will set $U=0$ without loss of generality. 
As another important observation, let us point out that contrary to the generic open system discussion in
the previous section, the kernel $W_{N}$ here gives rise to relaxation
and decoherence in a basis which is different from the eigenbasis
of the local dynamics $H$. This will render the dynamics much more complex, especially when including a transport detector, as we show in what follows.

However, before we continue with transport measurements and transport statistics,
let us briefly describe the system dynamics of $\rho$.
As already introduced in Sec. \ref{sec:integer_versus_fractional_JE}
the system dynamics is governed by the set of generally complex eigenvalues
$\left\{ \lambda_{n}\right\} $ of the superoperator $L$, Eq.~\eqref{eq:kernel_qu_sc_metal}, with the
corresponding eigenoperators. In absence of the parity drive, $\Gamma_{N}=0$,
the dynamics of the density matrix is given by the eigenmodes of $-i\left[H,\cdot\right]$
alone. The eigenoperators $\left|+\right\rangle \left\langle +\right|,\left|-\right\rangle \left\langle -\right|,\left|1_{\sigma}\right\rangle \left\langle 1_{\sigma}\right|$,
all have eigenvalues $0$, meaning that they correspond to the eigenstates
of $H$. The eigenoperators $\left|\pm\right\rangle \left\langle \mp\right|$
with the eigenvalues $-i\left(\epsilon_{\pm}-\epsilon_{\mp}\right)$
indicate the coherent dynamics. With finite $\Gamma_{N}$, the even
and odd subsectors couple. The eigenvalues are now $0,-\Gamma_{N},-2\Gamma_{N},-i\left(\epsilon_{\pm}-\epsilon_{\mp}\right)-\Gamma_{N}$,
where $-\Gamma_{N}$ is doubly degenerate, see Fig. \ref{fig:system}b.
These eigenvalues can be interpreted as the decay of physical quantities
as discussed in Refs. \cite{Splettstoesser_2010,Contreras_2012,Saptsov_2012}.
For this purpose, one needs to consider the structure of the corresponding
eigenoperators. While it is possible
to find a closed form for the eigenoperators for arbitrary system
parameters, the expressions are
quite cumbersome and thus not very instructive. We therefore consider the here relevant limit $\Gamma_{N}\ll\left|\epsilon_{+}-\epsilon_{-}\right|$,
where we may simplify the expressions considerably (done explicitly
in Appendix \ref{app:eigenmodes}). Namely, we find that the eigenoperators
belonging to $\lambda_{\pm}=-i\left(\epsilon_{\pm}-\epsilon_{\mp}\right)-\Gamma_{N}$
are still approximately given by $\left|\pm\right\rangle \left\langle \mp\right|$,
which now represent the coherent oscillations damped with the decoherence
rate $\Gamma_{N}$. The eigenvalue $\lambda_{0}=0$ corresponds to
the stationary state of the quantum master equation,
\begin{align}
\widehat{\rho}^{\text{st}} & \approx\frac{\left(1-\delta\right)^{2}}{4}\left|+\right\rangle \left\langle +\right|+\frac{1-\delta^{2}}{4}\sum_{\sigma}\left|1_{\sigma}\right\rangle \left\langle 1_{\sigma}\right| \nonumber\\\label{eq_rho_st}
 & +\frac{\left(1+\delta\right)^{2}}{4}\left|-\right\rangle \left\langle -\right|.
\end{align}
We observe that even though the parity switching rate is small, the
occupation of the odd state is of the same order as the even state
in $\rho^{\text{st}}$. This is simply due to the fact that while
parity switches from even to odd are rare, the same is true for the
reversed process from odd to even. Hence the system spends an approximately
equal amount of time in either parity sector. 
The eigenvalue $\lambda_{p}=-2\Gamma_{N}$ indicates the decay of
the fermion parity, given by the operator $\widehat{p}=e^{i\pi\widehat{n}},$
as discussed also in Ref. \cite{Saptsov_2012}. The doubly degenerate
eigenvalue $\lambda_{s,c}=-\Gamma_{N}$ relates to two
processes. For one, to the decay of spin, $\widehat{s}=\sum_{\sigma}\sigma\left|1_{\sigma}\right\rangle \left\langle 1_{\sigma}\right|$,
and for another, to the decay of what we refer to as the pseudo-charge
number $\widehat{c}=\left|+\right\rangle \left\langle +\right|-\left|-\right\rangle \left\langle -\right|$.
We baptize it in this way because in the absence of the
proximity effect, $E_{J}\rightarrow0$, we find $\widehat{c}\rightarrow\left|2\right\rangle \left\langle 2\right|-\left|0\right\rangle \left\langle 0\right|=\widehat{n}-1$. 

With respect to the Josephson effect, note that the normal metal itself
merely introduces relaxation and decoherence, but does not alter the
periodicity of the eigenspectrum with respect to $\phi$: the coherent
oscillations still occur with the frequency $\epsilon_{+}-\epsilon_{-}$,
which, as discussed above, are usually $2\pi$ periodic in $\phi$ (unless the system parameters are tuned to very special values). Hence, in the generic case of a spectrum with a minigap, the complex open system
spectrum has the same topology as the one shown in Fig. \ref{fig:open_FJE}e. This
will change now, when considering the combination of open system dynamics and transport measurements. Let us point out though, that while the transport measurement is indispensable, the presence of a \textit{nonequilibrium} drive due to the voltage-biased normal metal is equally important. For a pure equilibrium drive, the kernel $L$ would satisfy an equivalent of a PT symmetry, where braid phase transitions are forbidden even in the presence of a counting field~\cite{Ren_2012,Riwar_2019b}.

\section{Different flavors of full-counting statistics\label{sec:different_flavors_of_FCS}}

Generally, for a superconducting junction described by a Hamiltonian $H\left(\phi\right)$,
the operator for the supercurrent across the junction can be defined
as $I=2e\partial_{\phi}H\left(\phi\right)$. In Eq.~\eqref{eq:Josephson_potential_energy},
the phase bias is attached to the right contact, such that the operation
$\partial_{\phi}$ actually returns the current to the right,
\begin{equation}
I\equiv i eE_{JR}e^{i\phi}d_{\uparrow}^{\dagger}d_{\downarrow}^{\dagger}+\text{h.c.}=I_{SR}.
\label{eq:supercurrent_operator}
\end{equation}
By means of a simple unitary transformation, the Josephson
energy could be modified as $E_{JL}+E_{JR}e^{i\phi}\rightarrow E_{JL}e^{-i\phi}+E_{JR}$,
such that here, the current at the left interface would be measured.
In accordance with what we stated in the introduction and in Sec.~\ref{sec:integer_versus_fractional_JE}, the position of measurement is not a mere gauge choice, and gives rise to different predictions. Here, this difference is in particular due to the addition of a third (normal metal) contact, which injects an additional dissipative displacement current.
For the remainder of this paper we will stick for concreteness to the explicit example where the current is measured at the right contact (see also Fig.
\ref{fig:system}a). In order to map these results to the case where the detector is on the left, one has to mirror the entire device (that is exchange $E_{JL}\leftrightarrow E_{JR}$). Let us note that yet another physically distinct scenario would be to distribute the phase bias across both junctions with a factor $\zeta$, i.e., $E_{JL}+E_{JR}e^{i\phi}\rightarrow E_{JL}e^{-i\zeta\phi}+E_{JR}e^{i(1-\zeta)\phi}$. This would express the situation when a current detector couples to both currents at the left and right junction with this prefactor. Our results would certainly be sensitive to the value of $\zeta$. We disregard this option for simplicity, assuming that it is physically possible to build a current detector which couples only to the right contact.

\begin{figure}[htbp]
    \centering
    \includegraphics[width=0.8\columnwidth]{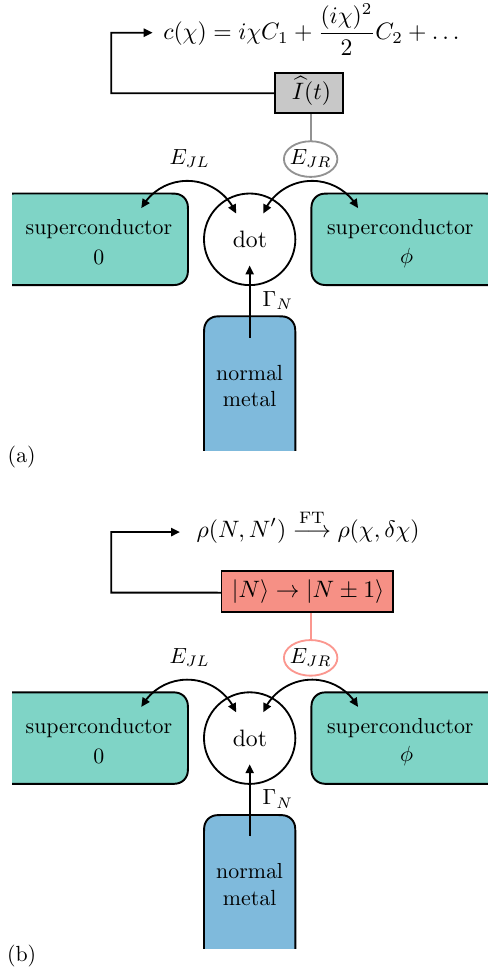}
    \caption{Different approaches to FCS. (a) The current at the right junction may be projectively measured at a given time $t$, and current measurements at different times can be correlated and integrated over time, in order to obtain the cumulants $C_k$. In between measurements, the system propagates freely. The cumulant generating function can then be reconstructed in a Taylor series. In this approach, the counting field $\chi$ is a fictitious quantity. (b) An ideal detector may be coupled at the right contact, such that for each transported Cooper pair, the detector changes its state $|N\rangle\rightarrow|N\pm 1\rangle$ where $N$ stands for the number of transported Cooper pairs. The detector thus continuously entangles with the system. As a consequence, the counting field $\chi$ is here an actual physical quantity: the detector momentum, related to $N$ by a Fourier transform (FT). However, after a projective measurement of the detector state, the information of the dissipation-free supercurrent is lost.}
    \label{fig:flavours_of_FCS}
\end{figure}

For starters, let us point out that in the absence of the normal metal, the even parity eigenstates $\left|\pm\right\rangle $
exhibit a dc Josephson effect,
\begin{equation}
\left\langle I\right\rangle _{\pm}=2e\partial_{\phi}\epsilon_{\pm}.\label{eq:Josephson_currents}
\end{equation}
In the odd parity sector, the system is ``poisoned'', and no supercurrent
flows, see also discussion after Eq.~\eqref{eq:spectrum_quantum_dot}. For the current expectation value, the main influence of the
normal metal is a reduction of the supercurrent in the stationary
state, due to the finite occupation of the poisoned state, see Eq.~\eqref{eq_rho_st}.

In the
following, we now want to go way beyond the current expectation value,
and describe the entire FCS of the transport, where the interplay
between quasiparticle-induced dissipation and current measurement
will give rise to a plethora of nontrivial effects. However, before
going down that road, we have to explicitly address the fact that there are several different
ways to define the FCS, which correspond to different measurement
schemes. While these differences do not play a role for purely dissipative
transport, in the presence of supercurrents, these different ``flavours''
of FCS give rise to markedly different results, and in particular
to different interpretations of the observed topological transitions.

\subsection{Averaging time-resolved current measurements}\label{sec:time-resolved_FCS}

In the context of superconducting transport, a straightforward access
to FCS is due to \cite{Romito_2004}, whereby the quantum master
equation is supplemented with a counting field $\chi$, $\dot{\rho}\left(\chi\right)=L\left(\chi,\phi\right)\rho\left(\chi\right)$,
such that
\begin{equation}
L\left(\chi,\phi\right)\cdot=-i\left[H\left(\phi-\chi\right)\cdot-\cdot H\left(\phi+\chi\right)\right]+W_{N}\cdot.\label{eq:L_chi}
\end{equation}
The cumulant generating function for the transported charges after
a measurement (integration) time $\tau$, $c\left(\chi,\tau\right)$,
is then computed via the moment generating function $m\left(\chi,\tau\right)$,
defined as
\begin{equation}
m\left(\chi,\phi,\tau\right)=\text{tr}\left[e^{L\left(\chi,\phi\right)\tau}\rho_{0}\right]\equiv e^{\tau c\left(\chi,\phi,\tau\right)},\label{eq:moment_generating_function}
\end{equation}
where $\rho_{0}$ is the initial state (which becomes irrelevant for
large measurement times $\tau$). Derivatives of the cumulant generating
function provide the cumulants $C_k$. For instance, the average current
is given as
\begin{equation}\label{eq_current_C1}
\langle I\rangle=C_1=\left.-ie\partial_{\chi}c\right|_{\chi\rightarrow0},
\end{equation}
and the current noise (usually denoted by the letter $S$) is given as
\begin{equation}\label{eq_noise_C2}
S=C_2=\left.\left(-i\right)^{2}e^{2}\partial_{\chi}^{2}c\right|_{\chi\rightarrow0},
\end{equation}
and so forth. In order to appreciate the difference to the other important notion of FCS (explained below), we have to go beyond this formal definition, and recapitulate in detail, \textit{how} the current
is actually measured in order to obtain the above cumulants. For this purpose, let us examine the first couple
of statistical moments a little more closely. The first moment (giving
rise to the current expectation value $I=C_1$) returns
\begin{align}
 & \left.-ie\partial_{\chi}m\right|_{\chi\rightarrow0}=\nonumber\\
 & -ie\text{tr}\left[\int_{0}^{\tau}dt_{1}e^{L\left(\phi\right)\left(\tau-t_{1}\right)}\left.\partial_{\chi}L\left(\phi\right)\right|_{\chi\rightarrow0}e^{L\left(\phi\right)t_{1}}\rho_{0}\right]
\end{align}
where
\begin{equation}
-ie\left.\partial_{\chi}L\right|_{\chi\rightarrow0}\cdot=e\left\{ \partial_{\phi}H\left(\phi\right),\cdot\right\} =\frac{1}{2}\left\{ I,\cdot\right\} ,
\label{eq:supercurrent_superoperator}
\end{equation}
with the anticommutator $\left\{ \cdot,\cdot\right\} $. As we see,
the FCS as defined above corresponds to a system evolving freely (that
is, without any detection event) for most of the time, and a projective
current measurement at a precise time step $t_{1}$, and subsequently,
integrating over all times $t_{1}$ from $0$ to a total measurement time $\tau$, as schematically represented in Fig.~\ref{fig:flavours_of_FCS}a. The zero-frequency limit of the FCS is when the measurement time approaches infinity, $\tau\rightarrow\infty$.
The picture becomes even more detailed, when going to the next moment,
providing the current-current correlations,
\begin{widetext}
\begin{align}
\left.\left(-i2e\right)^{2}\partial_{\chi}^{2}m\right|_{\chi\rightarrow0} & =\left(-i2e\right)^{2}\text{tr}\left[\int_{0}^{\tau}dt_{1}e^{L\left(\phi\right)\left(\tau-t_{1}\right)}\left.\partial_{\chi}^{2}L\left(\phi\right)\right|_{\chi\rightarrow0}e^{L\left(\phi\right)t_{1}}\rho_{0}\right]\nonumber\\
 & +2\text{tr}\left[\int_{0}^{\tau}dt_{1}\int_{0}^{t_{1}}dt_{2}e^{L\left(\phi\right)\left(\tau-t_{1}\right)}\left\{ I,\cdot\right\} e^{L\left(\phi\right)\left(t_{1}-t_{2}\right)}\left\{ I,\cdot\right\} e^{L\left(\phi\right)t_{2}}\rho_{0}\right].
\end{align}
\end{widetext}
While both the first and second line now indicate two
current measurements, the time difference between these two measurements
 is of the essence. While the second line accounts for projective
current measurements at times $t_{1}$ and $t_{2}$, which are sufficiently
far apart (with an unimpeded system evolution for the rest of the
time interval), the first line describes two current measurements
that occur within time intervals which are short with respect to the
superconductor correlation time $\Delta^{-1}$ (see also previous
section). For the interested reader, we refer to the diagrammatic
language for noise, which was first developed for time-independent
systems \cite{Thielmann_2003,Thielmann_2005,Thielmann_2005prb},
subsequently generalized to time-dependent systems \cite{Riwar_2013}
as well as finite frequency noise \cite{Dittmann_2018}. In this
language, measurements according to the first line are represented
by diagrams where the two current operators appear within the same
irreducible block.

At any rate, it is interesting to note that if the current detector
fails to measure time-resolved currents on a time-scale smaller than
$\Delta^{-1}$, the first line will be absent altogether. This can be understood as a high-frequency cut-off for the FCS. Such a deficient
detector would return a different moment generating function, given
as
\begin{equation}
m_{\text{cut-off}}\left(\chi,\phi,\tau\right)=\text{tr}\left[e^{L_{\text{cut-off}}\left(\chi,\phi\right)\tau}\rho_{0}\right],
\end{equation}
with
\begin{equation}
L_{\text{cut-off}}\left(\chi,\phi\right)=L\left(\phi\right)+i\frac{\chi}{2e}\left\{ I,\cdot\right\} ,\label{eq:complete_kernel_small_chi}
\end{equation}
which is nothing but a first order in $\chi$ approximation of the
full $L\left(\phi,\chi\right)$. Thus, while the eigenspectrum of
$L_{\text{cut-off}}\left(\phi,\chi\right)$ asymptotically approaches
the one for the full $L\left(\phi,\chi\right)$ for low values of
$\chi$ (low cumulants), the global properties (arbitrarily high cumulants)
differ decisively. In particular, while the full $L$ is $2\pi$-periodic
in $\chi$, $L\left(\phi,\chi+2\pi\right)=L\left(\phi,\chi\right)$,
reflecting the fact that the detector measures the supercurrent in integer portions
of Cooper pairs, this information is lost in $L_{\text{cut-off}}$.
Likewise, we cannot in general hope to see the same topological phase
transitions along $\chi$ for the two scenarios. We conclude that
if we are interested in understanding and measuring the global properties
of the FCS with respect to $\chi$ (relevant for fractional charges
as defined in \cite{Riwar_2019b}) by means of projective current
measurements, a current detector which can resolve beyond the time-scale
$\Delta^{-1}$ is required.

\subsection{Continuous entanglement with a charge transport detector}

There is a different approach to FCS, whereby an explicit detector
is included in the model description of the system \cite{Levitov_1996,Schaller_2009,Pluecker_2017,Pluecker_2017prb},
keeping track of the number of charges exchanged at a given interface.
Hence, the counting field $\chi$ is here not merely an auxiliary mathematical
object without any physical meaning. To the contrary, it has a well-defined
precise interpretation: $\chi$ is the detector momentum \cite{Pluecker_2017,Pluecker_2017prb}.
Because of this, the global properties defined in $\chi$-space are
much more tangible compared to the notion described in the above
section, where large $\chi$ can only be reached by measuring a sufficiently
high number of cumulants. Here, an analysis (read-out) of the detector
state may directly provide the moment generating function for finite $\chi$, in contrast to the previously introduced approach, where $m$ as a function of $\chi$ would have to be reconstructed essentially by analytic continuation, starting from $\chi=0$.

Following the lines of \cite{Schaller_2009,Pluecker_2017,Pluecker_2017prb},
a detector measuring transport at the interface to the right superconductor
can be modeled by supplementing the proximity Hamiltonian $H_{J}$
with the detector degrees of freedom $\left|N\right\rangle $ indicating
the number $N$ of measured Cooper pair transport events,
\begin{equation}
\rightarrow H_{J}=\frac{1}{2}\left(E_{JL}+E_{JR}e^{i\phi}\sum_{N}\left|N-1\right\rangle \left\langle N\right|\right)d_{\uparrow}^{\dagger}d_{\downarrow}^{\dagger}+\text{h.c.}
\end{equation}
Thus, the detector state changes as $|N\rangle\rightarrow|N\pm 1\rangle$, for each Cooper pair leaving or entering the right contact, see also Fig.~\ref{fig:flavours_of_FCS}b. Note that the detector itself is ideal in the sense that it does not
have any internal dynamics apart from this coupling (i.e., the Hamiltonian of the isolated detector is zero). The quantum system plus
detector have a much larger state space, described by the density
matrix $\rho_{S+D}=\sum_{N,N'}\rho\left(N,N'\right)\otimes\left|N\right\rangle \left\langle N'\right|$,
illustrating the fact that the detector will be entangled
with the system during the measurement. Similar to the previous notion of FCS, $\tau$ here stands for the total measurement time. Whereas in Sec.~\ref{sec:time-resolved_FCS} $\tau$ represented the total time interval over which the current measurements should be averaged, here $\tau$ stands for the total time elapsed since the coupling to the detector (and thus the build up of entanglement) started. Also for continuous entanglement, a related zero-frequency FCS can be defined, by analyzing the asymptotic behaviour for $\tau\rightarrow\infty$.

The additional detector degree of freedom can be compactified by a double Fourier
transform,
\begin{equation}
\rho\left(\chi,\delta\chi\right)=\sum_{N}\sum_{N'}e^{i\chi\left(N+N'\right)}e^{i\delta\chi\left(N-N'\right)}\rho\left(N,N'\right),
\end{equation}
resulting in the quantum master equation
\begin{equation}
    \dot{\rho}\left(\chi,\delta\chi\right)=L_{\text{CE}}\left(\chi,\delta\chi,\phi\right)\rho\left(\chi,\delta\chi\right) \ .
\end{equation}
We note that the Lindbladian describing the continuous entanglement with
the detector, $L_{\text{CE}}$, is related to the above, first version
of FCS, described by $L\left(\chi,\phi\right)$ in Eq.~(\ref{eq:L_chi}),
as
\begin{equation}\label{eq_LCE_from_L}
L_{\text{CE}}\left(\chi,\delta\chi,\phi\right)=L\left(\chi,\phi-\delta\chi\right).
\end{equation}
The variables $\chi$ and $\delta\chi$ can be thought of as the classical
and quantum component of the detector momentum. As we see, the classical
detector momentum corresponds to the counting field introduced in
the FCS of Ref. \cite{Romito_2004}.

The quantum part $\delta\chi$ on the other hand simply enters as
a shift in the superconducting phase, and may therefore at first sight
seem innocuous. It is however this shift, which makes all the difference.
Namely, for the continuously entangling detector, the moment generating function is defined as the Fourier
transform of a projective measurement of the detector state in its eigenbasis $\left|N\right\rangle $,
\begin{align}
m_{\text{CE}}\left(\chi,\tau\right) & =\sum_{N}e^{i2N\chi}\text{tr}\left[\rho\left(N,N,\tau\right)\right]\nonumber\\
 & =\int_{0}^{2\pi}\frac{d\delta\chi}{2\pi}\text{tr}\left[\rho\left(\chi,\delta\chi,\tau\right)\right],
\end{align}
where for the second identity, we used the fact that the integration
over $\delta\chi$ results in the projection onto the diagonal elements
$\rho\left(N,N'\right)\rightarrow\rho\left(N,N\right)$. Importantly,
due to $\delta\chi$ appearing as a shift in $\phi$, we can relate
this moment generating function to the first one, Eq.~(\ref{eq:moment_generating_function}),
as follows
\begin{equation}
m_{\text{CE}}\left(\chi,\tau\right)=\int_{0}^{2\pi}\frac{d\phi}{2\pi}m\left(\chi,\phi,\tau\right).\label{eq:relationship_between_different_FCS}
\end{equation}
Overall, we note that since $L_\text{CE}$ and likewise $m_\text{CE}$, Eqs.~\eqref{eq_LCE_from_L} and~\eqref{eq:relationship_between_different_FCS}, can be constructed from $L$ and $m$ respectively, Eqs.~\eqref{eq:L_chi} and~\eqref{eq:moment_generating_function}, we consider $L$ to be the more fundamental construction of FCS. Therefore, it will suffice to analyse the topological properties of $L(\chi,\phi)$.

However, the above phase shift $\delta\chi$ plays an important role when it comes to analyzing the topological eigenspectrum of $L$, due to the presence of supercurrents. If supercurrents were absent, there would be no phase-dependence
of the transport $m\left(\chi,\phi\right)=m\left(\chi\right)$, such
that $m$ and $m_{\text{CE}}$ are equivalent. However, for supercurrents
being present, the two notions of FCS differ, in that the supercurrents
are averaged out in $m_{\text{CE}}$. One can convince oneself of
this fact, simply by means of the Josephson relations given in Eq.
(\ref{eq:Josephson_currents}), where $\int_{0}^{2\pi}d\phi I_{S\pm}\sim\epsilon_{\pm}\left(2\pi\right)-\epsilon_{\pm}\left(0\right)$
must be zero, due to the $2\pi$-periodicity of the eigenspectrum
$\epsilon_{\pm}$ in $\phi$.
This cancellation of the supercurrent is a consequence of the detector always
being ideally coupled to the interface at which it counts the number of transported Cooper pairs. It thus entangles with the coherent
transport (and the entanglement continuously increases as the measurement
goes on), such that when projectively reading out the detector state,
the information about the supercurrent is destroyed. Nonetheless,
such a continuously entangled detector may serve for an understanding
of the topology of the dissipative part of transport. Moreover, as
we will show below, such a detector will give rise to a novel transport phase,
which can be interpreted as a statistical mix between a fractional
and a trivial transport.

Let us conclude this section by pointing out the following. In this work, we aim at understanding the topological properties of the eigenspectrum of $L$ along both the $\chi$ and $\phi$ coordinates. While this is endeavour is formally well-defined, thanks to Eq.~\eqref{eq:L_chi}, from a more practical point of view, we see that both of the above flavours of FCS come with their advantages and disadvantages. Ultimately we have the choice between measuring individual cumulants without destroying the supercurrent information (in accordance with the construction of $L$), which however allows us to only explore the vicinity of $\chi\approx 0$ (since the measurement of arbitrarily high cumulants is experimentally challenging), or, via $L_\text{CE}$, explore the full $\chi$-space (since the detector and thus $\chi$ are here physical) but at the expense of losing the supercurrent information, and thus losing the $\phi$-dependence. Moreover, realistically, a detector measuring the charge that arrived at one of the superconducting contacts most likely involves supplementing said contact with a capacitance, which thus renders the detector nonideal (its Hamiltonian is no longer zero). This leads us to consider below a third variation to obtain information about the transport statistics: continuous weak measurement. However, this approach is very challenging to cast into a general form, which is why we first present our results for the topology of $L$, and identify a particularly interesting topological regime, for which we formulate a specifically tailored version of weak transport measurement.

\section{Fractional charge versus fractional Josephson effect\label{sec:fractional_charge_vs_fractional_JE}}

We have so far established a framework to describe the open system dynamics
of a superconductor-normal metal hybrid circuit, including the FCS, based on the Lindbladian $L\left(\chi,\phi\right)$ in Eq.~(\ref{eq:L_chi}). Let us now explore the topological properties of the eigenspectrum of $L\left(\chi,\phi\right)$, $\{\lambda_n\left(\chi,\phi\right)\}$. In order to analyze the topology of the eigenspectrum, keeping track of the eigenvalue labelling will be important. In Sec.~\ref{sec:normal_metal}, we have already introduced the labelling $\{\lambda_0,\lambda_\pm,\lambda_p,\lambda_{s,c}\}$, motivated by the physical interpretation of the decay processes of the corresponding eigenmodes. When including the counting field $\chi$, the eigenspectrum will be modified, $\lambda_n(\phi)\rightarrow\lambda_n(\chi,\phi)$. For finite $\chi$ we will still use the same labelling of indices, which is however somewhat tricky because of braid phase transitions, whereby certain eigenvalues swap places. We therefore use the convention, that the labelling $n$ shall be done according to Sec.~\ref{sec:normal_metal} at the reference point $\chi=0$.

As detailed in the previous section, the counting field and the superconducting phase difference
appear per se as independent parameters. Similarly to Refs. \cite{Ren_2012,Li_2013,Riwar_2019b},
it turns out that the nonequilibrium drive via the normal metal will give rise to topological transitions
in the spectrum $\lambda_n$. However, in Refs. \cite{Ren_2012,Li_2013,Riwar_2019b} only the counting field $\chi$ was considered as a relevant coordinate. Here, we have the 2D space $(\chi,\phi)$. In 2D, exceptional points appear, as we will explain in more detail below. When considering cuts of the complex spectrum along either $\chi$ or $\phi$, these exceptional points generate a braid phase transition, and a resulting broken periodicity in either $\chi$ or $\phi$. We will interpret phases with a broken periodicity along $\chi$ as a transport with fractional
charges (along the lines of Ref. \cite{Riwar_2019b}).
Intriguingly, including here the superconducting phase $\phi$, our theory
also predicts braid phase transitions in $\phi$. There are several
nontrivial phases, which can be classified as a fractional Josephson
effect, in the sense of having a spectrum with broken periodicity
in $\phi$. We will refine this statement in the following. Importantly, since transitions along both coordinates appear due to exceptional points in the 2D space $(\chi,\phi)$, we conclude that the
fractional charge defined in the transport statistics ($\chi$) and
the fractional Josephson effect ($\phi$)
are intimately related, but distinct concepts in a generic open system context.

\subsection{Braid phase transitions due to exceptional points in $\left(\chi,\phi\right)$}

\begin{figure}[htbp]
\centering\includegraphics[width=1\columnwidth]{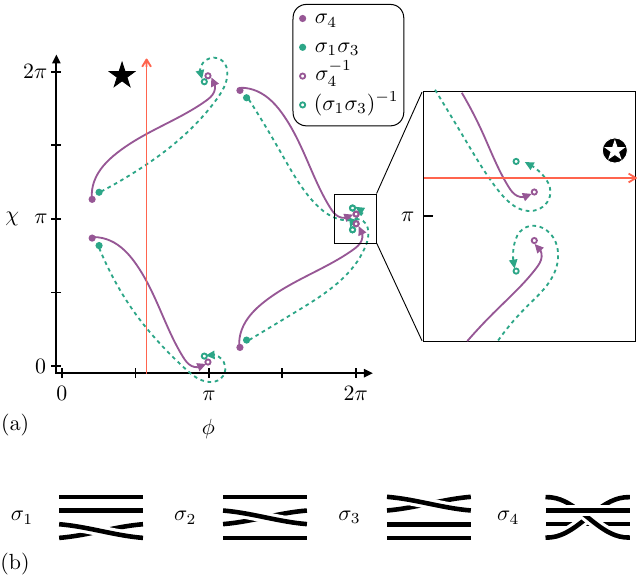}

\caption{(a) Positions of exceptional points in $\left(\chi,\phi\right)$-space
for $E_{JL}=E_{JR}\equiv E_{J}$, $\epsilon=0.1E_{J}$, and $\Gamma_{N}=0.5E_{J}$.
To each exceptional point, one may assign braid generators (similar
to a topological charge), marked with solid and empty circles. Since
four eigenvalues partake in braid phase transitions, we need the braid
group for four strands, given in b). In fact, this braid group is
complete already with the first three generators, $\sigma_{1}$, $\sigma_{2}$,
and $\sigma_{3}$. The fourth braid generator is only added for convenience; it can be expressed as
$\sigma_{4}=\sigma_{1}^{-1}\sigma_{3}^{-1}\sigma_{2}\sigma_{1}\sigma_{3}$.
Due to overall ``braid charge'' conservation, each exceptional point
must have its negative counter part, to which the inverse braid generator
is assigned. Such a pair of exceptional points is connected via an
arrow (solid purple and dashed green). The braids along a particular
axis (either $\chi$ or $\phi$, see two examples marked with red
arrows), see subsequent figures, can be constructed from a) by the
following rule. To know the topology (i.e., the braid word) of a spectrum
along a given path, one needs to assemble all the generators of the
connection lines of two exceptional points, in the order they are
crossed. For instance, the red arrow along $\chi$, gives rise to
the braid word $\left(\sigma_{1}\sigma_{3}\sigma_{4}\right)^{2}=\left(\sigma_{2}\sigma_{1}\sigma_{3}\right)^{2}$,
which corresponds to the topological phase given in Fig. \ref{fig:braids_chi}d.
The red line along $\phi$ {[}see inset of a){]}, returns the braid
word $\sigma_{1}\sigma_{3}\sigma_{4}\sigma_{1}^{-1}\sigma_{3}^{-1}=\sigma_{2}$,
and thus the topological phase from Fig. \ref{fig:braids_phi}d. \label{fig:exceptional_points}}

\end{figure}

In order to describe the topological
phase transitions in the spectrum of $L(\chi,\phi)$, we first have to establish some technical details
regarding braid theory. As already stated, in general, the eigenspectrum
of $L\left(\chi,\phi\right)$, $\{\lambda_n\left(\chi,\phi\right)\}$, is complex. Considering the space spanned
by $\left(\chi,\phi\right)$, we can think of the eigenspectrum as
a complex band in a 2D Brillouin zone (where $\chi$ and $\phi$ are
coordinates of the 2D torus). Therefore, the touching of two complex
bands, $\lambda_n=\lambda_{m\neq n}$, leads to two independent conditions, which may be satisfied
for particular values of both $\chi$ and $\phi$. That is, a touching of two complex bands occurs in isolated points on the 2D torus. In Fig. \ref{fig:exceptional_points}a, we show the location of exceptional points for a chosen parameter set. As it turns out, band degeneracy points
can occur for typical system parameters (see caption of Fig.~\ref{fig:exceptional_points}). Locally, at the degeneracy
points, the two eigenvalues partaking in the degeneracy can be described
by a complex square root function $\pm\sqrt{z}$ (where $z\sim\chi+i\phi$).
In the literature of topological transitions in open quantum systems,
these touching points are commonly referred to as exceptional points, see, e.g., Ref.~\cite{Bergholtz_2021} (and references therein).

When choosing a closed path in $\left(\chi,\phi\right)$ around an
exceptional point, the two eigenvalues which touch at the exceptional
points, perform a braid. Thus, to each of the exceptional points,
one may assign generators of the braid group; the braid generator
may be considered as a generalization of the notion of a topological
charge carried by a degeneracy point (see, e.g., \cite{Riwar_2019b}). Given a certain ordering of the eigenvalues, the index $j$ of the braid generator $\sigma_{j}$ (see Fig. \ref{fig:exceptional_points}b) indicates, which two eigenvalues perform a braid. We here choose the order of the labels as ${\lambda_0,\lambda_+,\lambda_-,\lambda_p}$, see, e.g., Fig.~\ref{fig:braids_chi}. For instance, the braid generator $\sigma_1$ thus braids $\lambda_0$ with $\lambda_+$, $\sigma_2$ braides $\lambda_+$ with $\lambda_-$ and finally $\sigma_3$ braids $\lambda_-$ with $\lambda_p$. Note that the eigenvalues $\lambda_{s,c}$ are inert, in the sense that they depend neither on $\chi$ nor $\phi$, and do not partake in braiding. This is why we do not have to include them for the analysis of the topology of the eigenspectrum. While $\partial_{\chi}\lambda_{s}=0$ can be understood
by the complete symmetry of the system with respect to spin, $\partial_{\chi}\lambda_{\overline{n}}=0$ stems from the effective elimination of the many-body interactions
within the dissipative (quasiparticle-induced) processes, due to $\mu\gg U$. At any rate, we only have to consider braid generators with four strands. For four strands, the set of three braid generators, $\sigma_{1}$,
$\sigma_{2}$, and $\sigma_{3}$, see Fig. \ref{fig:exceptional_points}b, is complete and describes the whole braid group. For convenience, we have furthermore introduced a braid generator to directly braid the first and fourth strand ($\lambda_0$ and $\lambda_p$), $\sigma_{4}=\sigma_{1}^{-1}\sigma_{3}^{-1}\sigma_{2}\sigma_{1}\sigma_{3}$. This generator is however non-fundamental in the sense that it can be constructed out of $\sigma_{1,2,3}$.

Due to the $2\pi$-periodicity of $L$ in $\chi$ and $\phi$, there
must be an overall conservation of the ``braid charge'' (similar
to \cite{Riwar_2019b}, where this was discussed for complex $\chi$).
As a consequence, to each exceptional point with a given braid generator
$\sigma_{j}$ (see Fig. \ref{fig:exceptional_points}b), there must
exist a partner point, with the inverse braid generator. These two
partner points are connected by a line, see \ref{fig:exceptional_points}a.
In fact, since the exceptional points are locally described by the
square root function, these lines can be understood as the corresponding
branch cuts, with the two partner points as origin points for the
branch cut. When considering the spectrum along one particular parameter
(either $\chi$ or $\phi$, see red arrows in Fig. \ref{fig:exceptional_points}a),
the braid word for the spectrum can be constructed as follows. One
simply has to add for each branch cut which is crossed, the corresponding
braid generator in the order it is crossed. In order to know the chirality
of the braid generator (i.e., whether one has to add a given $\sigma_{j}$
or $\sigma_{j}^{-1}$) one may follow a ``right hand rule'': take the cross product of the tangential vector indicating the path taken (red arrow in Fig.~\ref{fig:exceptional_points}a) and the tangential vector of the branch cut, at the given point, where these two lines cross. The direction of the cross product vector decides the chirality. In Fig.~\ref{fig:exceptional_points}a, we show two examples of paths (as red arrows), one along $\chi$ for a fixed value of $\phi$, and conversely one along $\phi$ with fixed $\chi$. Along these paths, the topological phases discussed below emerge (marked with a star and an inverse star symbol, cf Secs.~\ref{sec:fractional_charges} and~\ref{sec:fractional_JE}).

At this point, let us comment on the importance of the fact that the dissipative processes due to $W_N$ relax into a basis different from the eigenbasis of $H$. If the coupling to the environment would be such that the dissipative processes occurred in the basis of $H$, as is the case, e.g. in Eq.~\eqref{eq_Lindblad_generic}, then the addition of the counting field $\chi$ would not give rise to any interesting topological transitions. Here, the eigenvalues $\lambda_\pm(\phi)=\pm i[\epsilon_+(\phi)-\epsilon_-(\phi)]-(\Gamma_++\Gamma_-)/2$ would simply receive a $\chi$-dependence as $\lambda_\pm(\chi,\phi)=\mp i[\epsilon_+(\phi+\chi)-\epsilon_-(\phi-\chi)]-(\Gamma_++\Gamma_-)/2$. Due to $\epsilon_\pm$ being gapped for the quantum dot circuit, the complex spectrum eigenspectrum would here be trivial for all values of $(\chi,\phi)$. The normal metal providing an out-of-equilibrium electron source thus plays an essential role as the driver of topological phase transitions. We note that usually, processes which change the parity of superconducting circuits are considered detrimental (referred to as quasiparticle poisoning, see, e.g., Refs.~\cite{Lutchyn_2005,Shaw_2008,Catelani_2011,Leppakangas_2012,Fu_2009,vanHeck_2011,Rainis_2012,Goldstein_2011,Budich_2012,Pekola2013}). Here, we provide a rare counter example, where they are at the origin of an interesting effect.

\subsection{Fractional charges}\label{sec:fractional_charges}

\begin{figure}[htbp]
\centering\includegraphics[width=1\columnwidth]{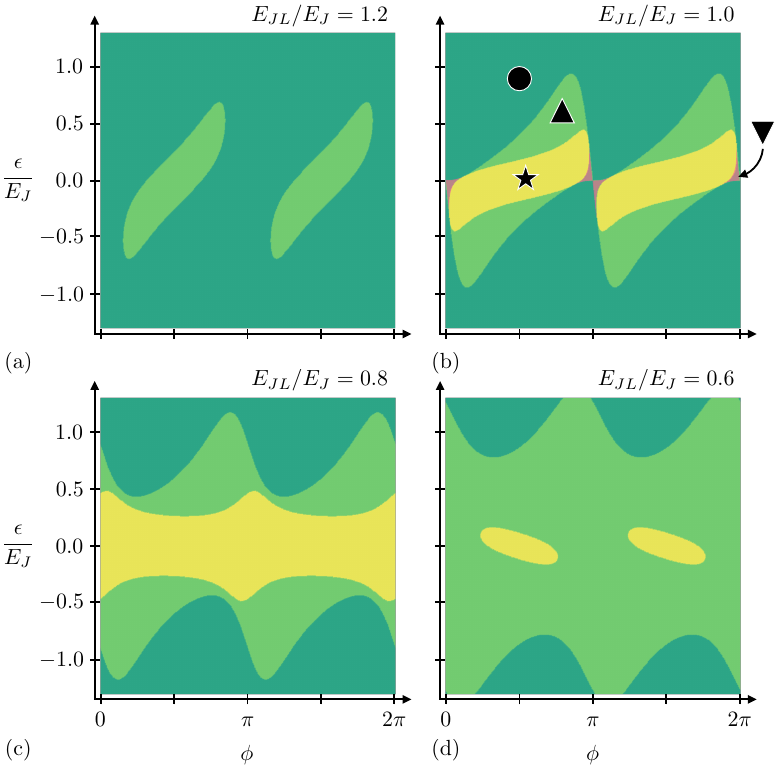}
\caption{Map of the different topological phases of the eigenspectrum $\lambda_n\left(\chi\right)$ (with $n=\{0,+,-,p\}$), as a function of the detuning $\epsilon$, and the phase bias $\phi$,
for different asymmetries of the Josephson energies, $E_{JL}/E_{J}$
(a-d). In (b) we mark all four possible
topological phases with the symbols of circle, triangle,
inverted triangle, and star. Out of those, only the
yellow phase (star) is a topological phase with fractional charge
$e^{*}=e/2$. For asymmetric junctions, $E_{JL}/E_{J}=0.8$ (see panel c), this phase is connected
for all $\phi$ for a certain interval of $\epsilon$ close to $0$.\label{fig:map_chi}}
\end{figure}

\begin{figure}[htbp]
\centering\includegraphics[width=1\columnwidth]{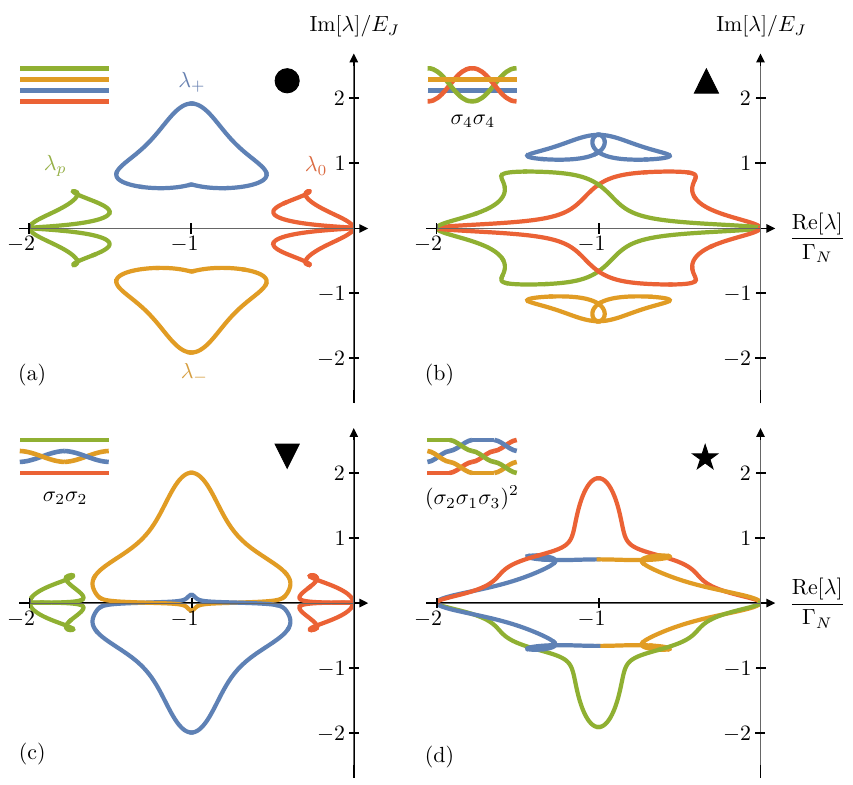}
\caption{The real and imaginary parts of the spectrum $ \lambda_n\left(\chi\right)$ (with $n=\{0,+,-,p\}$),
drawn parametrically for $\chi=\left[0,2\pi\right)$, for the four
different topological phases mapped out in Fig. \ref{fig:map_chi}
(a-d), and the definitions for the generators of the braid group,
$\sigma_{1,2,3}$ (e). The corresponding inverse generators $\sigma_{1,2,3}^{-1}$
can be constructed by braiding with opposite chirality. There is a
trivial phase (a), where all complex eigenvalues form separate bands.
The corresponding braid word is trivial. In (b) the eigenvalues $\lambda_{0}$
and $\lambda_{p}$ braid twice, such that the total spectrum remains
$2\pi$-periodic in $\chi$. This spectrum can be described by the
braid word $\sigma_{4}\sigma_{4}$ with $\sigma_{4}=\sigma_{1}^{-1}\sigma_{3}^{-1}\sigma_{2}\sigma_{3}\sigma_{1}$.
In (c) the same double braiding occurs but with the eigenvalues $\lambda_{+}$
and $\lambda_{-}$, characterized by the braid word $\sigma_{2}\sigma_{2}$.
Finally, in (d) the braid leads to eigenvalues $\lambda_{0}$ and
$\lambda_{p}$, respectively $\lambda_{+}$ and $\lambda_{-}$, swapping
places. Here, the eigenspectrum is $4\pi$-periodic in $\chi,$ thus
breaking the $2\pi$-periodicity of $L\left(\chi\right)$. Here, the
transport can be described by eigenmodes with fractional charge $e^{*}=e/2$.
\label{fig:braids_chi}}
\end{figure}

Let us now analyze explicitly the plethora of braid phase transitions of the eigenspectrum of $L\left(\chi,\phi\right)$ along $\chi$ with fixed $\phi$. While we could in principle use the information of the exceptional points in $(\chi,\phi)$-space, as discussed above, we note that for explicit calculations, there is a mathematically more efficient approach, which was discussed in Ref. \cite{Riwar_2019b} (and further detailed in Appendix~\ref{app:excep_points}). Namely, the trick is to describe braid transitions in $L\left(\chi\right)$ (for fixed $\phi$)
by generalizing to complex counting fields $e^{i\chi}\rightarrow z\in\mathbb{C}$
(and $e^{-i\chi}\rightarrow1/z$), such that the real counting fields
are represented on the unit circle, $\left|z\right|=1$. The exceptional points now do no longer appear in the 2D space of $(\chi,\phi)$, but in the complex 2D space of $z$. A braid phase
transition in the $\chi$-space occurs when an exceptional point traverses
the unit circle. Therefore, in order to keep track of the topological
phases, we simply have to compute the number of exceptional points
residing within the unit circle. Based on the definition for $L$
in Eq.~(\ref{eq:L_chi}), we find that the positions of the exceptional points can be obtained analytically by means of the quartic equations,
\begin{equation}\label{eq_EP_quartic}
    \sum_{i=0}^4 p_i z^i =0 \quad\text{and}\quad \sum_{i=0}^4 q_i z^i =0 \ ,
\end{equation}
where the coefficients $p_i$ and $q_i$ depend on all the system parameters and $\phi$. Their explicit forms are given in Appendix~\ref{app:excep_points}. Note that both equations have to be fulfilled individually, such that there are two sets of roots for $z$, one for the first, and one for the second polynomial equation. As explained above, for a given root $z_0$, one simply has to test if $|z_0|\lessgtr 1$ and count the total number of roots inside the unit circle, which enables us to draw maps of the topological phases as in Fig.~\ref{fig:map_chi} as a function of all the system parameters and $\phi$. We find overall four different types of braids for the eigenspectrum along $\chi$, which are labelled in Fig.~\ref{fig:map_chi}b with the sphere, triangle, upside down triangle, and star symbols, and explicitly drawn at example points in parameter space in Fig.~\ref{fig:braids_chi}. In the upper left corner of each panel in Fig.~\ref{fig:braids_chi}, we also provide the braid word describing the topology of the spectrum.

There is a trivial phase, shown in Fig.~\ref{fig:braids_chi}a (sphere symbol). Here, the eigenvalues $\{\lambda_0,\lambda_+,\lambda_p,\lambda_-\}$ do not swap places within the entire interval $\chi\in[0,2\pi)$. There are two topological phases (triangle, and upside-down triangle), shown in Figs.~\ref{fig:braids_chi}b and c. In Fig.~\ref{fig:braids_chi}b, the eigenvalues related to the stationary state $\lambda_0$ and the parity decay $\lambda_p$ perform a braid. However, this braid does not break the $2\pi$-periodicity in $\chi$, as they braid twice, as indicated by the braid words $\sigma_4 \sigma_4$ and $\sigma_2 \sigma_2$.

Finally, there is a topological phase where the $2\pi$-periodicity in $\chi$ is broken, see Fig.~\ref{fig:braids_chi}d. After progression of $\chi$ by $2\pi$ the eigenvalues $\lambda_0$ and $\lambda_p$, as well as $\lambda_+$ and $\lambda_-$ have swapped places, leading to an overall $4\pi$-periodicity of the spectrum. Along the lines of Ref.~\cite{Riwar_2019b}, such a spectrum can be interpreted as transporting a charge with half the unit as compared to the $2\pi$-periodic phases. For the sake of completeness, let us reiterate the arguments of Ref.~\cite{Riwar_2019b}. First of all, note that as per the two definitions of the FCS, Eqs.~\eqref{eq:L_chi} and~\eqref{eq_LCE_from_L}, the charge is counted in units of Cooper pairs, with charge $2e$. The trivial phases with a spectrum $2\pi$-periodic in $\chi$ therefore transport charges in units of $2e$. The in the $4\pi$-periodic phase, the transported charge is $e^*=2e/2=e$. Charge quantization is broken in the sense that physically, the s-wave superconducting contacts can only accept integer Cooper pairs (due to the large $\Delta$ limit). The non-equilibrium drive due to the normal-metal induces the topological phase of Fig.~\ref{fig:braids_chi}d, where the contacts seem to accept half-integer Cooper pairs. In fact, this breaking of charge quantization seems already to some extent indicative of a fractional Josephson effect, which we will discuss in detail in moment.

According to Ref.~\cite{Riwar_2019b}, there are two important ways to define fractional charges in $\chi$. Let us first consider the zero-frequency limit of FCS. As already mentioned above, the measurement time $\tau$ is here to be taken as infinite, $\tau\rightarrow\infty$. Consequently, in the transport statistics, only the eigenvalue with the least negative real part, $\lambda_0$ is visible (see also Ref.~\cite{Bagrets_2002,Bagrets_2003}), as can be seen when considering the definition of the moment generating function $m$ in Eq.~\eqref{eq:moment_generating_function}. As $\tau$ increases, all higher eigenmodes $\lambda_{n\neq 0}$ become exponentially suppressed. In fact, the cumulant generating function in this limit can be computed simply as
\begin{equation}
    \lim_{\tau\rightarrow \infty}c(\chi,\tau)=\lambda_0(\chi)\ .
\end{equation}
That is, for a hypothetical experimenter measuring the true zero-frequency FCS, the information of the higher modes would be lost. However, something nontrivial remains.
Suppose that we were able to measure a sufficiently large number of cumulants $C_k$ to reconstruct $c(\chi)$, and thus $m(\chi)$, for finite values of $\chi$. This would essentially correspond to analyzing the eigenvalue $\lambda(\chi)$ first close to $\chi\approx 0$ and then analytically continuing to finite $\chi$. If the cumulants $C_k$ are measured up to a sufficiently high (ideally infinitely high) order $k$, the cumulants could thus be used to reconstruct the periodicity of $\lambda_0$ in $\chi$ and thus determine unit of the charge being transported. Curiously, when the zero mode $\lambda_0$ partakes in a braid phase transision, the analytic continuation would clearly provide a $4\pi$-periodic moment generating function, indicating transport in units of $e$, in spite of the system physically transporting charges into the superconducting reservoirs in units of $2e$.

The interpretation of the broken periodicity as a fractional charge works also for finite measurement times $\tau$, when the transport statistics still depend on $\tau$, and the decaying modes $\lambda_{n\neq 0}$ are still detectable. Here, Ref.~\cite{Riwar_2019b} argues, that the spectrum consisting of complex bands with broken periodicity in $\chi$ can be exactly mapped to a fictitious open quantum system which transports charges in units given by the periodicity in $\chi$. In the topological phase shown in Fig.~\ref{fig:braids_chi}d, both $\lambda_0$ and $\lambda_p$ as well as $\lambda_+$ and $\lambda_-$ merge into two complex bands, each with periodicity $4\pi$. Thus, this corresponds to two fictitious bands transporting charge $e$ instead of $2e$.

There is an additional final point to be discussed. Namely, the spectrum as shown in Fig.~\ref{fig:braids_chi} can only be measured when adopting the first version of FCS, outlined in Sec.~\ref{sec:time-resolved_FCS}, i.e., when projectively measuring the current at local times, and correlating them to obtain the transport statistics. If the FCS is instead measured with a continuously entangling detector, according to Eq.~\eqref{eq:relationship_between_different_FCS} the moment generating function is integrated over $\phi$. Since the moment generating function is a regular expectation value, the integral over $\phi$ (including the normalization prefactor $1/2\pi$) can be understood effectively as a statistical average over a homogeneous probability distribution in $\phi$. It is in this sense, that the transport statistics obtained via $m_\text{CE}$ are to be understood as a statistical mix of different topological phases. For instance, in Fig.~\ref{fig:map_chi} for a given $\epsilon$ one can draw a horizontal line along $\phi$, and thus evaluate how many distinct topological phase regions the line crosses. In particular, there is thus the possibility to observe a statistical mix of topological phases with different transported charge units, either the charge $2e$ for the phases in Figs.~\ref{fig:braids_chi}a, b, and c or the charge $e$ for Fig.~\ref{fig:braids_chi}d. Such an effect was not posssible in Ref.~\cite{Riwar_2019b}, where only normal metal contacts were considered, and thus the FCS was $\phi$-independent.

\subsection{Fractional Josephson effect}\label{sec:fractional_JE}

As we have seen just now, one particular topological phase along $\chi$ (Fig.~\ref{fig:braids_chi}d) indicated transport with a charge $e$ instead of $2e$ which would be the default charge of the superconducting contact. Here, we want to analyse the topological properties of the eigenspectrum along $\phi$ for different values of $\chi$. To this end, we proceed similarly as above, this time, by replacing $e^{i\phi}\rightarrow\widetilde{z}$ (and $e^{-i\phi}\rightarrow1/\widetilde{z}$) and analyzing the positions of exceptional points in the space of general, complex $\widetilde{z}$. Also here, this position can be evaluated again by means of a quartic equation with the same form as in Eq.~\eqref{eq_EP_quartic}, with $z\rightarrow\widetilde{z}$ and the new coefficients $p_i,q_i\rightarrow\widetilde{p}_i,\widetilde{q}_i$, depending on $\chi$ instead of $\phi$. Again, their explicit form is given in Appendix~\ref{app:excep_points}.

\begin{figure}[htbp]
\centering\includegraphics[width=1\columnwidth]{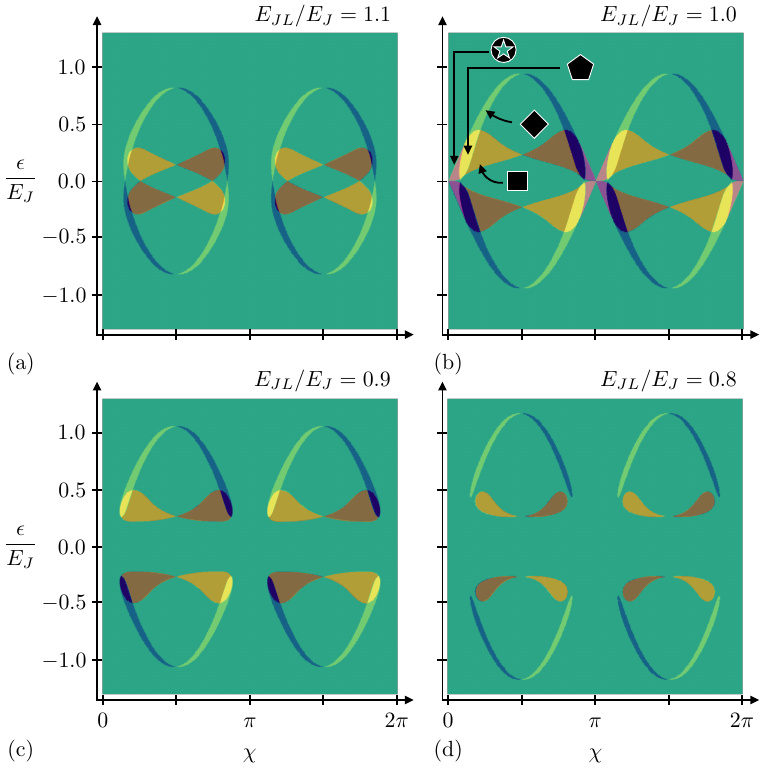}
\caption{Map of the different topological phases of the spectrum $ \lambda_n\left(\phi\right)$, as a function of the detuning $\epsilon$, and the counting field
$\chi$, for different asymmetries of the Josephson energies, $E_{JL}/E_{J}$
(a-d). Apart from the trivial phase, there are here four distinct
topological phases, marked in (b) with the symbols of square, rhombus,
pentagon, and inverted star. For each topological phase at a given
$\left(\epsilon,\chi\right)$, there is a partner phase at $\left(-\epsilon,\chi\right)$
which can be obtained through complex conjugation of the bands $\lambda_{x}\rightarrow\lambda_{x}^{*}$.
\label{fig:map_phi}}
\end{figure}

\begin{figure}[htbp]
\centering\includegraphics[width=1\columnwidth]{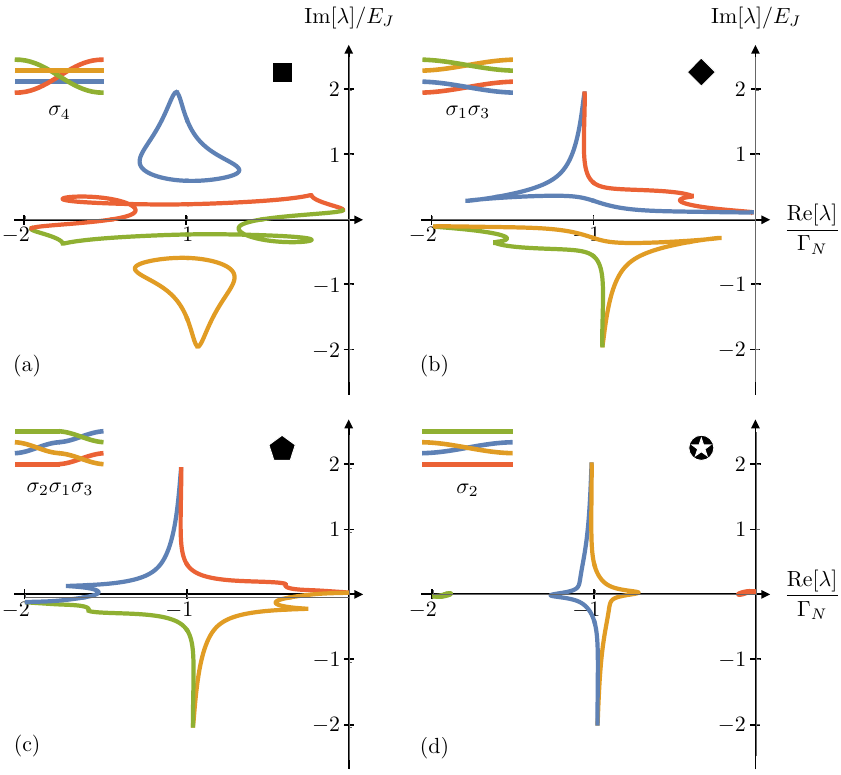}
\caption{The real and imaginary parts of the spectrum $ \lambda_n\left(\phi\right)$,
drawn parametrically for $\phi=\left[0,2\pi\right)$, for the four
different topological phases mapped out in Fig. \ref{fig:map_phi}
(a-d). In all phases, the $2\pi$-periodicity in $\phi$ is broken.
In (a) and (d), two eigenvalues participate in a braid, either $\lambda_{0}$
and $\lambda_{p}$ in (a), or $\lambda_{+}$ and $\lambda_{-}$ in
(d). In (b), both $\lambda_{0}$ and $\lambda_{+}$ as well as $\lambda_{p}$
and $\lambda_{-}$ exchange places during a $2\pi$-sweep of $\phi$.
In (c), all eigenvalues interchange, leading to an $8\pi$-periodic
phase. The phase depicted in (d) has a mapping to a closed system
fractional Josephson effect including weak dissipation. The other
phases (a-c) do not have such a correspondence, since they involve
the eigenvalues $\lambda_{0}$ and $\lambda_{p}$ (see also main text).
\label{fig:braids_phi}}
\end{figure}

The resulting map of topological phases is shown in Fig.~\ref{fig:map_phi}. Here, there are overall five different phases to be observed, a trivial one, and four topologically nontrivial ones, denoted by the square, diamond, pentagon and inverted star symbols (as indicated in Fig.~\ref{fig:map_phi}b). Here, all of the nontrivial phases break $2\pi$-periodicity along $\phi$. Hence, in this broad sense, all of these phases may be interpreted as a fractional Josephson effect. In particular, apart from the $4\pi$-periodic phases in Fig.~\ref{fig:braids_phi}a and b (denoted with the square and diamond symbols), there is in fact an $8\pi$-periodic phase (pentagon symbol), where all four non-inert eigenvalues partake in a braid. Here, we can think of the interaction with the magnetic field generating the phase bias $\phi$ in terms of a charge $e/2$, similar to parafermionic circuits~\cite{Zhang_2014,Orth_2015}. Note that a fractional charge $e/2$ could not be observed in the topological properties along $\chi$ discussed previously. We can therefore see this as a nice example illustrating why the topological properties along $\chi$ and along $\phi$ should in general be considered distinct effects. Topological transitions along both parameters are related due to their common generation by means of the exceptional points in $(\chi,\phi)$, as shown in Fig.~\ref{fig:exceptional_points}, however, as it turns out, their configuration may be such that different braid phases occur along the two parameters. 

In addition, there is a topological phase (inverted star symbol in Fig.~\ref{fig:map_phi}b) which deserves the label of a fractional Josephson effect in a more narrow sense. Namely, the complex spectrum here (shown in Fig.~\ref{fig:braids_phi}d) can be continuously mapped to the open system spectrum of the actual Majorana-fermion circuit, shown in Fig.~\ref{fig:open_FJE}f. That is, the shape of the eigenvalues in Fig.~\ref{fig:braids_phi}d permits the explicit interpretation of the spectrum as two eigenvalues related to the coherent (Hamiltonian) dynamics $\lambda_\pm$ which have now a closed minigap, and remaining standard eigenvalues related to decay and stationary state, which do not partake in the braid. Importantly, we note that for a closed system, the gap in the Josephson spectrum can only be closed when using at least four superconducting contacts (and thus three phase differences $\phi_{1,2,3}$) as shown in Ref.~\cite{Riwar2016}. As already pointed out in Sec.~\ref{sec:integer_versus_fractional_JE}, for the circuit considered here, with only two superconducting contacts (and the single phase difference $\phi$), the closed system cannot stabilize a closing of the minigap: any deviation from $\epsilon=0$ or $E_{JL}=E_{JR}$ opens a gap. Here we show, that a gap closing can be stabilized by means of the interplay between a nonequilibrium drive (due to the normal metal) and a measurement of the transport statistics (nonzero $\chi$).

Let us conclude this section by summarizing, that for a generic open quantum circuit, the emergence of fractional charges defined in the FCS ($\chi$) and a fractional Josephson effect, indicating the unit of charge with which the magnetic field interacts ($\phi$) are in so far related, as they are generated by exceptional points in the 2D space spanned by $(\chi,\phi)$. They are however also distinct in the sense that these exceptional points produce different braids when analyzing the spectrum either along the $\chi$-space (where either $2\pi$- or $4\pi$-periodic spectra emerge) or along the $\phi$-space (where we find $2\pi$-, $4\pi$- and even $8\pi$-periodic spectra). Moreover, we can show that the fractional Josephson effect, and thus a closing of the minigap in the Josephson energy (imaginary part of the complex eigenspectrum) can be stabilized when combining nonequilibrium and transport measurements, a feature which is impossible for a closed (dissipation-free) circuit.

Finally, let us explicitly point out what we have already indicated at the beginning of Sec.~\ref{sec:different_flavors_of_FCS}. Namely, there is a left-right asymmetry in the occurrence of topological phases, as can be seen when swapping $E_{JL}\leftrightarrow E_{JR}$, e.g., when comparing Figs.~\ref{fig:map_chi}a and c as well as Figs.~\ref{fig:map_phi}a and c. This is due to the fact, that the current is measured asymmetrically (at the right contact), as we have discussed when defining the current operator, Eq.~\eqref{eq:supercurrent_operator}. Exact left-right symmetry is only achieved by mirroring the circuit \textit{and} the detector placement.

\section{Finite counting field as weak measurement \label{sec:Weak_measurement}}

\begin{figure}[htbp]
\includegraphics[width=0.8\columnwidth]{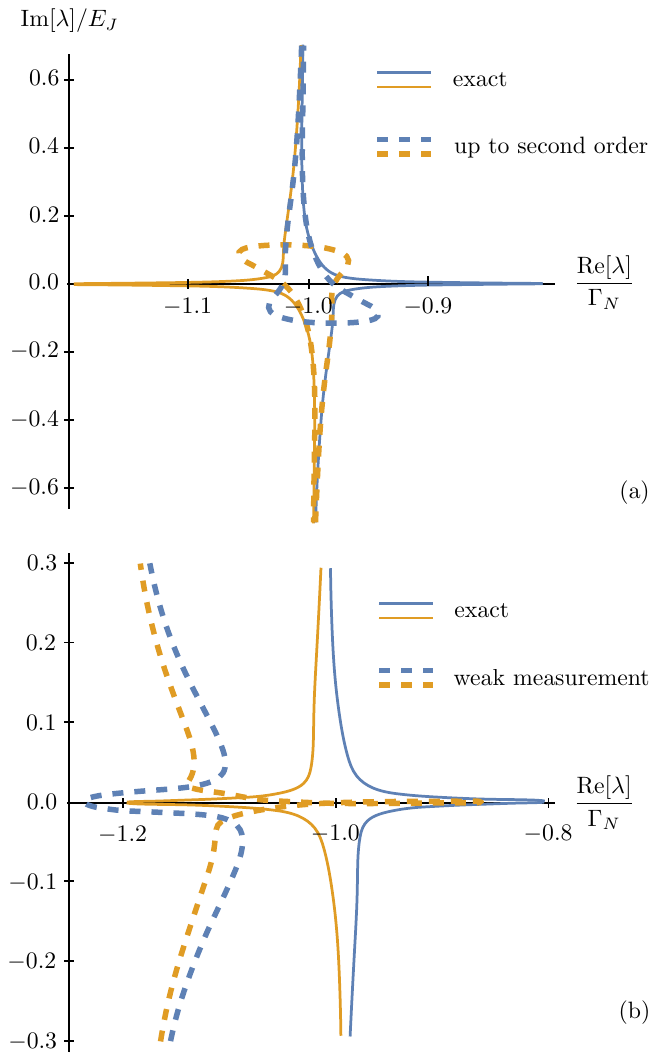}
\caption{(a) The dashed lines are analytically continued eigenvalues up to second
order, while the solid lines are eigenvalues with a finite counting field (see also Fig.~\ref{fig:braids_phi}d).
The analytically continued eigenvalues match asymptotically only away from $\phi=\pi$. Near $\phi=\pi$ however, they do not perform the same braiding as in Fig.~\ref{fig:braids_phi}d, such that the topological phase cannot be observed. (b) The solid lines are again eigenvalues for ideal detector kernel with finite counting field (Fig.~\ref{fig:braids_phi}d),
and the dashed lines are eigenvalues for the weak measurement kernel.
The eigenvalues do not match up exactly (due to a distortion, see main text), but they exhibit the same braiding. These plots are for the following parameter values:
$\frac{\epsilon}{E_{JR}}=0.007$, $\frac{E_{JL}}{E_{JR}}=1.0$, $\frac{\Gamma_{N}}{E_{JR}}=0.2$,
$\chi=-0.04$, $\frac{\omega_{m}\delta r}{E_{JR}}=0.02$ and $\xi=\pi/2$.}
\label{fig:analytic_cont}
\end{figure}

While we have learned above that a trivial circuit with a quantum dot coupled to superconducting and normal metal contacts provide an unexpected wealth of open system topological phase transitions, there are some important remaining caveats especially with regard to the nature of the detector. In particular, the observation of the topological phase transitions along $\phi$-space (see Figs.~\ref{fig:braids_phi}) are in fact virtually impossible, when adhering to the idealized detection schemes depicted in Fig.~\ref{fig:flavours_of_FCS}. As for the detection scheme with time-local projective current measurements, Fig.~\ref{fig:flavours_of_FCS}a, the finite $\chi$ parameter regime can only be approached, by measuring an increasing number of cumulants $C_k$ of the supercurrent (and, in fact, by including finite measurement times $\tau$ in order to extract all the eigenmodes, see Appendix~\ref{app:analytic_continuation}) and then analytically continuing the eigenmodes $\lambda_n(\chi)$ starting from the extracted $\partial_\chi^{k}\lambda_n(0)$. It goes without saying that, such a procedure is in and of itself extremely challenging expermentally. Moreover, note also, that there is no convergence if we aim to go across a topological phase transition. Let us explicitly illustrated this fact with the example of the topological spectrum shown in Fig.~\ref{fig:braids_phi}d. For $\chi=0$ (while keeping all other parameters the same) the spectrum is trivial.
The analytically
continued eigenvalues, starting at $\chi=0$, are defined as
\begin{equation}
\begin{split}
\lambda_{i}^{\text{ac}}(\chi,\phi)=\lambda_{i}(0,\phi)+\chi\left.\partial_{\chi}\lambda_{i}(\chi,\phi)\right|_{\chi=0}\\+\frac{\chi^{2}}{2!}\left.\partial_{\chi}^{2}\lambda_{i}(\chi,\phi)\right|_{\chi=0}+\cdots
\end{split}
\end{equation}
Now we can compare the analytically continued eigenspectrum to the exact one (without expansion around $\chi=0$)
to see if they still braid in the same way, keeping all other parameters same. In Fig.~\ref{fig:analytic_cont}a we compare the parametric plots of the analytically continued eigenvalues to second order and exact eigenvalues for finite counting field, and can easily conclude that analytically continued eigenvalues do not reproduce the same braid. We find that the $4\pi$-periodic fractional Josephson effect only emerges when going to arbitrary high order cumulants, which is an outright prohibitive requirement from an experimental viewpoint.

This issue could be avoided if $\chi$ was a real, physical parameter, which it is not when utilizing time-local current correlations (Fig.~\ref{fig:flavours_of_FCS}a). It would be, if instead an explicit physical detector was present (Fig.~\ref{fig:flavours_of_FCS}a), however, here there is the aforementioned problem, that the continuous entanglement between system and detector destroys the information of the supercurrent.
This prompts us to study alternative measurement schemes, where the transport measurement satisfies both the requirements of the counting field being physical, and at the same time preserving information about the supercurrent. As it turns out, these requirements can be met by a weak continuous measurement of the current.

Weak measurement has been studied extensively in several contexts (for an instructive review, see Ref.~\cite{Clerk_2010}). A weak continuous measurement of the current could for instance be envisaged along similar lines as in Ref.~\cite{Liu_2010}, where it was proposed to weakly measure spins via an incident polarized photon beam and exploiting the Faraday effect. Due to the magnetic field emitted by the supercurrent, it is in principle perceivable to use a similar setup here to obtain information about the transport. However, in the light of massive experimental advances in the interaction and control of superconducting circuits with transmission lines~\cite{Wallraff_2004,Gu_2017,Wendin_2017}, we deem it informative to briefly sketch an ``all-circuit'' realization of the weak measurement (i.e., without relying on polarized photon beams). For this purpose we study coherent scattering in a nearby SQUID inductively coupled to the circuit, loosely inspired by Ref.~\cite{Steinbach_2001}.
In the following we will provide a highly simplified model of a SQUID detector and its interaction with the circuit, as a proof of principle for a weak measurement of the supercurrent, and show how it can be used to simulate a counting field. Finally, we will investigate as an example, how the topological phase from Fig.~\ref{fig:braids_phi}d can be probed with this setup.

\subsection{SQUID detector for weak current measurement}

\begin{figure}[htbp]
\begin{centering}
\includegraphics[width=0.7\columnwidth]{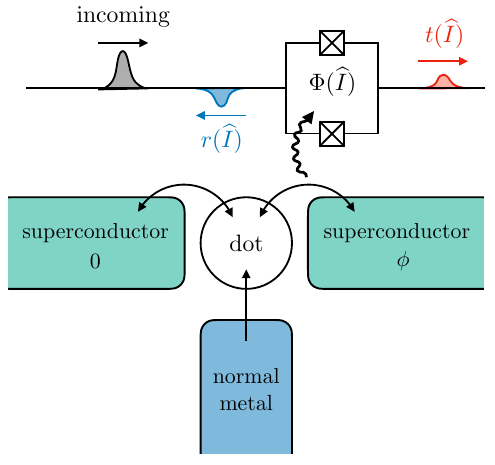}
\par\end{centering}
\caption{Setup for weak measurement by means of a SQUID detector. An incoming wave packet will be scattered at the SQUID. The inductive coupling between the main circuit and the SQUID shall be tailored such that the outgoing scattered state depends on the supercurrent entering the superconductor on the right. A subsequent projective measurement of the scattered state realizes a form of weak measurement of the supercurrent.}\label{fig:SQUID_detector_setup}
\end{figure}

The SQUID detector consists of two superconducting lines connected via two Josephson junctions (with Josephson energy $E_{J,\text{SQUID}}$) in parallel, see Fig.~\ref{fig:SQUID_detector_setup}. The weak measurement is then implemented by means of the following points. (i) The current from the main circuit produces a magnetic field that can
interact with the SQUID. The interaction strength can be estimated based on Amp\`ere's law, see, e.g., Ref.~\cite{Riwar_2021}. (ii) If we send a signal from one end of the detector it will be reflected and transmitted at the SQUID. (iii) The reflection and transmission coefficients are sensitive to the flux enclosed by the SQUID and therefore depend on the current from
the quantum dot.

As we will develop now, a subsequent evaluation of the scattered state will provide us with classical information about the supercurrent without completely suppressing it. In particular, we will show that the measurement is weak because of a highly reflective nature of the SQUID (i.e., full reflection is the default event, without obtaining any information about the current), and continuous in the sense that there is a repeated initiation of incoming waves after a given time interval, the inverse of which represents the detection frequency.

In order to describe the scattering problem, let us start by writing down the Hamiltonian for the SQUID detector. Overall it is composed of three parts, $H_{\text{SQUID}} =H_{L}+H_{R}+V$ describing the left (right) conductor line $H_L$ ($H_R$) and the SQUID part connecting the two, $V$. Each of these subparts can be written as
\begin{align}
H_{L} & =\frac{1}{2C_{0}}\sum_{j=-\infty}^{0}q_{j}^{2}+\frac{1}{2L_{0}}\sum_{j=-\infty}^{-1}\left(\frac{\varphi_{j+1}-\varphi_{j}}{2e}\right)^{2}\label{eq:SQUID_hamiltonian_HL} \\
H_{R} & =\frac{1}{2C_{0}}\sum_{j=1}^{\infty}q_{j}^{2}+\frac{1}{2L_{0}}\sum_{j=1}^{\infty}\left(\frac{\varphi_{j+1}-\varphi_{j}}{2e}\right)^{2} \label{eq:SQUID_hamiltonian_HR}\\
V & =-\frac{\gamma}{2e^2}\cos\left(\varphi_{1}-\varphi_{0}\right)\approx \gamma\left(\frac{\varphi_1-\varphi_0}{2e}\right)^2+\text{const.}\ , \label{eq:SQUID_hamiltonian_V}
\end{align}
where we chose for convenience a discrete lattice representation of the conductor lines~\cite{Pozar_2012}, which are characterized by the capacitances $C_0$ and the inductances $L_0$. The charge and phase variables $q_j$ and $\varphi_j$ on the lattice nodes $j$ satisfy the commutation relations $[q_j,\varphi_{j'}]=i2e\delta_{jj'}$. The interaction with the circuit is included via the coupling prefactor in $V$ (which is chosen to  have the units of an inverse inductance). When tuning the externally applied flux to half flux quantum, we get
\begin{equation}\label{eq_gamma_def}
\gamma =2e^2\lambda E_{J,\text{SQUID}} I\ ,
\end{equation}
where $I$ is the current operator, as defined in Eq.~\eqref{eq:supercurrent_operator}. The coupling constant $\lambda$ can be estimated from Amp\`ere's law (as already mentioned, see Ref.~\cite{Riwar_2021}).  Note that for simplicity, we assume that the signals will have low amplitudes in $\varphi$, such that we may expand $V$ up to quadratic order, see Eq.~\eqref{eq:SQUID_hamiltonian_V}. Note that the tuning to half-flux makes the SQUID highly reflective; for $I=0$, the SQUID can be considered as a hard wall from the point of view of an incident wave, and only weakly transmittive for a finite $I$ since the local inductance of the SQUID is very high.

Let us briefly touch on an important point regarding the spatial resolution of the current measurement. Namely, depending on the circuit geometry and detector placement, the SQUID could in principle couple to both the left and right supercurrents (and thus fails to reproduce the topological phase transitions discussed above, and in general complicate the discussion). In the most generic case, we would thus actually have a nonzero $\zeta$ parameter, describing this nonideal coupling, see the discussion in Sec.~\ref{sec:different_flavors_of_FCS} after Eq.~\eqref{eq:supercurrent_operator}. In order to avoid such subtleties, we assume that the SQUID is placed more towards the right contact than the actual quantum dot. This is fine, because it is plausible to assume that the Cooper pairs, once they enter the right contact, are distributed very fast (according to the group velocity of the Nambu-Goldstone mode within the superconductor bulk, see, e.g., Ref.~\cite{Altland_Simons_book}). Hence, the bottleneck current is the tunneling current between dot and superconducting contact, which can be still observed deep within the right contact, neglecting the high-frequency displacement currents inside the bulk.

To continue, we note that the individual Hamiltonians $H_{L,R}$ each have a linear dispersion relation, for bosonic modes propagating in 1D, $E_k\approx \omega_0 |k|$ (valid for $|k|\ll 1$), with $\omega_0=1/\sqrt{L_0C_0}$ and the unitless wave vector $k$. We then assume that one creates an incoming signal at a certain energy $E$ with wave vector $k_E=E/\omega_0$, carried by the conductor lines, which can scatter at the SQUID. The transmission amplitude can be computed by means of the Fisher-Lee formula~\cite{Fisher_and_Lee_1981},
\begin{equation}
t^{\text{L}}(E)=-iv_{E}\lim_{j\to\infty,j'\to-\infty}G_{jj'}^{+}(E)e^{-ik_{E}(j-j')}\label{eq:Fisher_Lee_formula}
\end{equation}
where $v_{E}=\partial_{k}E_{k}|_{E_{k}=E}$
and $G_{jj'}^{+}(E)$ is the retarded single-particle (here, single-boson) Green's function. We obtain
\begin{equation}
t^{\text{L}}(E)=-i\frac{\gamma L_{0}}{k_{E}}\label{eq:transmission_coeff}
\end{equation}
The explicit calculation is shown in Appendix
\ref{app:transmission_coeff}. This final result is valid up to first order in $\gamma$ (in accordance with the assumption that tunneling is weak, $|t|\ll 1$). Importantly, for half flux $\Phi_{\text{ext}}=\Phi_0/2$, which will be our default parameter setting from now on, the transmission coefficient is directly proportional to the current $I$, see Eq.~\eqref{eq_gamma_def}. In particular, it changes sign if the current changes sign. Due to left-right
symmetry of SQUID detector we can easily deduce that
\begin{align}
t^{\text{L}}(E) & =t^{\text{R}}(E)=t(E)\nonumber\\
r^{\text{L}}(E) & =r^{\text{R}}(E)=r(E)
\end{align}
Thus, the reflection and transmission coefficients can be cast into a standard scattering matrix
\begin{equation}
S=\left(\begin{array}{cc}
r^{\text{L}} & t^{\text{R}}\\
t^{\text{L}} & r^{\text{R}}
\end{array}\right)=\left(\begin{array}{cc}
r & t\\
t & r
\end{array}\right)
\end{equation}
Since the scattering matrix is unitary $\left(S^{\dagger}S=\mathbb{I}\right)$,
we can deduce two equations that give us a relation between the reflection
and transmission coefficients. The first equation is $r^{*}t+t^{*}r=0$ which leads us to conclude that the reflection coefficient must be real since the transmission coefficient in Eq.~\eqref{eq:transmission_coeff} is imaginary.
The other equation, conservation of probability $|t|^{2}+|r|^{2}=1$,
lets us calculate the amplitude of $r$.

As already stated above the SQUID is weakly transmitting, therefore
most of the signal will be reflected,
\begin{equation}
r=1-\delta r\label{eq:reflection_coeff}
\end{equation}
with $\delta r\ll 1$. From this, we derive
\begin{equation}
|t|^{2}=2\delta r\label{eq:transmission_reflection_coeff}
\end{equation}
upto first order in $\delta r$. These identities will help us now in constructing the Master equation including the influence of the SQUID detector.

\subsection{Master equation including weak measurement}

Now we are ready to develop the master equation for the quantum dot
system in presence of the SQUID detector and show how it simulates a
counting field. To write down the master equation we need Kraus operators
that describe the weak measurement process. 

For simplicity, we assume that the scattering time of the wave packet at the SQUID is very short, much shorter than the dynamics due to the coupling with the superconductors (defined by the energy scale $\sim E_{JL,JR}$) and the coupling to the normal metal ($\sim\Gamma_N$). Thus, we are entitled to treat the different parts of the system dynamics independently, and add them up for the final Master equation.

Keeping therefore the circuit state constant, let the initial state of the circuit plus detector (no normal metal) before
scattering at the SQUID be the factorized state
\begin{equation}
\begin{split}
\left|\Psi_{\text{in}}\right\rangle  =\left[\vphantom{\frac12} C_{-}\left|I_{-}\right\rangle  +C_{0\uparrow}\left|I_{0\uparrow}\right\rangle  +C_{0\downarrow}\left|I_{0\downarrow}\right\rangle\right.   \\ \left. +C_{+}\left|I_{+}\right\rangle \vphantom{\frac12}\right]\otimes \left|\text{in,R}\right\rangle 
\end{split}
\end{equation}
where the states $\left|I_{\mp}\right\rangle ,\left|I_{0\sigma}\right\rangle$ ($\sigma=\uparrow,\downarrow$)
are the eigenvectors of the current operator $I$ and $C_{\mp},C_{0\sigma}$
are complex-valued wave function amplitudes, satisfying the normalization condition $\left|C_{-}\right|^{2}+\left|C_{0\uparrow}\right|^{2}+\left|C_{0\downarrow}\right|^{2}+\left|C_{+}\right|^{2}=1$. The vector $\left|\text{in,R}\right\rangle $
represents the normalized state that depicts the incoming signal, without loss of generality assumed to originate from the
right end of the conductor line. For the sake of completeness, let us provide the explicit forms of the current operator eigenbasis, in terms of the quantum dot charge basis, $|0\rangle, |1_\sigma\rangle, |2\rangle$, as introduced below Eq.~\eqref{eq:quantum_dot_hamiltonian}. The eigenstates with nonzero eigenvalues $I_\mp=\mp e E_{JR}$ are given as $|I_\mp\rangle=\left(|0\rangle\pm ie^{i\phi}|2\rangle\right)/\sqrt{2}$, and the degenerate pair of zero eigenvalues $I_{0\sigma}=0$ belong to the eigenvectors $|I_{0\sigma}\rangle=|1_\sigma\rangle$.

After scattering, the factorized initial state gets weakly entangled, resulting in the final state
\begin{equation}\label{eq_final_state1}
\begin{split}
\left|\Psi_\text{f}\right\rangle =\left[\vphantom{\frac12} t_{-} C_{-}\left|I_{-}\right\rangle +t_{+}C_{+}\left|I_{+}\right\rangle  \vphantom{\frac12}\right]\otimes\left|\text{out,L}\right\rangle\\ 
+\left[\vphantom{\frac12} r_{-} C_{-}\left|I_{-}\right\rangle  +\sum_\sigma C_{0\sigma}\left|I_{0\sigma}\right\rangle +r_{+} C_{+}\left|I_{+}\right\rangle \vphantom{\frac12}\right]\otimes \left|\text{out,R}\right\rangle \ ,
\end{split}
\end{equation}
where $r_{\pm}\text{ and }t_{\pm}$ are reflection and transmission
coefficients, respectively, corresponding to eigenvalues $I_{\pm}$. We have furthermore made use of the fact, that for $I_{0\sigma}=0$, the signal gets completely reflected (due to the half-flux tuning).

In fact, the above final state shown in Eq.~\eqref{eq_final_state1} is meaningful, if the experimenter is merely measuring the presence or absence of a transmitted wave. We note however, that an additional important piece of information can be extracted from the scattered state: the aforementioned sensitivity of the transmission amplitude on the sign of the current, $t_-=-t_+$. In terms of the outgoing signal, this sign change can be understood as a $\pi$-phase shift, which could in principle be detected by an appropriate interference setup. Then, we have three instead of two detection outcomes, which should therefore be cast into the final wave function
\begin{equation}\label{eq_final_state2}
\begin{split}
\left|\Psi_\text{f}'\right\rangle =t_{-} C_{-}\left|I_{-}\right\rangle \otimes\left|\text{out,L$-$}\right\rangle +t_{+}C_{+}\left|I_{+}\right\rangle \otimes\left|\text{out,L$+$}\right\rangle\\ +\left[\vphantom{\frac12} r_{-} C_{-}\left|I_{-}\right\rangle  +\sum_\sigma C_{0\sigma}\left|I_{0\sigma}\right\rangle +r_{+} C_{+}\left|I_{+}\right\rangle \vphantom{\frac12}\right]\otimes \left|\text{out,R}\right\rangle  \ ,
\end{split}
\end{equation}
where the states $\left|\text{out,L$\mp$}\right\rangle$ represent a measurement of a transmitted wave (outgoing to the left) including a determination of its relative phase shift with respect to the initial wave, leading to the extra index $\mp$.

Depending on the two possible detection scenarios, we have either two or three possible measurement outcomes for the ancilla
system, denoted by \text{the index $q\in\{0,1\}$, or $q\in\{-,0,+\}$.} The projection onto the different measurement outcomes is described in the second scenario by the three Kraus operators
\begin{align}
M_{\mp} & =t_{\mp}\left|I_{\mp}\right\rangle \left\langle I_{\mp}\right|\\
M_{0} & =r_{-}\left|I_{-}\right\rangle \left\langle I_{-}\right|+\sum_\sigma\left|I_{0\sigma}\right\rangle \left\langle I_{0\sigma}\right| +r_{+}\left|I_{+}\right\rangle \left\langle I_{+}\right|
\end{align}
For the first scenario, the $\mp$ outcomes are merged into a single Kraus operator $M_1=M_-+M_+$. Independent of the specific measurement basis, it is easy to  check that the requirement $\sum_{q}M_{q}^{\dagger}M_{q}=\mathbf{1}$
is satisfied. We notice, that due to the highly reflective nature of the SQUID, we may use Eq.~\eqref{eq:reflection_coeff}, to express the Kraus operator
\begin{equation}
    M_0=\mathbf{1}+\delta M_0\ ,
\end{equation}
where $\delta M_0$ scales linear in $\delta r$, and thus quadratic in $t$, see also Eq.~\eqref{eq:transmission_reflection_coeff}. The Kraus operators $M_\mp$ on the other hand scale linearly in $t$. This different scaling behaviour is important now for the derivation of the Master equation including weak measurement

We now assume that there is a repetition of incoming signals according to a measurement frequency $\omega_m$, that is, the weak entanglement and subsequent projective measurement occurs on average every time interval $\sim 1/\omega_m$. The time evolution of the density matrix due to this process (still neglecting the influence from the superconducting and normal metal contacts) can then be given as
\begin{equation}
    \dot{\rho}=\omega_m \left(\sum_q M_q \rho M_q^\dagger -\rho \right)\ .
\end{equation}
The right-hand side can be expanded up to second order in $t$, resulting in the Master equation
\begin{equation}
    \dot{\rho}=\omega_m \left(K_0+K_-+K_+\right)\rho\ ,
\end{equation}
with the definitions of the superoperators $K_\mp\cdot=M_\mp\cdot M_\mp$ and $K_0\cdot=\delta M_0\cdot +\cdot \delta M_0$.

These superoperators can be reexpressed using the quantum dot creation and annihilation operators as follows, 
\begin{align}
K_{0}\cdot&=-2\delta r\left(1+d_{\uparrow}^{\dagger}d_{\downarrow}^{\dagger}d_{\downarrow}d_{\uparrow}\right)\cdot\left(1+d_{\uparrow}^{\dagger}d_{\downarrow}^{\dagger}d_{\downarrow}d_{\uparrow}\right)\nonumber \\
K_{-}\cdot &=A\cdot A-\frac{i}{2}\left\{ \delta re^{i\phi}d_{\uparrow}^{\dagger}d_{\downarrow}^{\dagger}-\delta re^{-i\phi}d_{\downarrow}d_{\uparrow},\cdot\right\} \nonumber \\
K_{+}\cdot &=A\cdot A+\frac{i}{2}\left\{ \delta re^{i\phi}d_{\uparrow}^{\dagger}d_{\downarrow}^{\dagger}-\delta re^{-i\phi}d_{\downarrow}d_{\uparrow},\cdot\right\} \label{eq:weak_measurement_operators}
\end{align}
where $A$ is a Hermitian operator given by
\begin{equation}
    A=\sqrt{\delta r}\left(1+ie^{i\phi}d_{\uparrow}^{\dagger}d_{\downarrow}^{\dagger}-ie^{-i\phi}d_{\downarrow}d_{\uparrow}+d_{\uparrow}^{\dagger}d_{\downarrow}^{\dagger}d_{\downarrow}d_{\uparrow}\right)\ .
\end{equation}
In order to derive the above $K$ superoperators, we have used the  property $t_{-1} =-t_{1}$, which further implies $\left|t_{-1}\right|^{2}  =\left|t_{1}\right|^{2}=\left|t\right|^{2}$ and $\delta r_{-1}  =\delta r_{1}=\delta r$. We notice that $K_\pm$ are actually of the form $K_\pm = A\cdot A\mp \delta r/(2 e E_{JR}) \{I,\cdot\}$ with the current operator $I$ as defined in Eq.~\eqref{eq:supercurrent_operator}. This observation will be of great use in a moment.

Let us now reintroduce the dynamics due to the superconducting and normal metal contacts, captured by the Lindbladian $L(\phi)$, see Eq.~\eqref{eq:kernel_qu_sc_metal}. In addition, we keep a register $n$ which stores the classical information of the outcome of the above described weak continuous measurement, $\rho\rightarrow \rho(n)$. The full master equation can be written as
\begin{align}
\dot{\rho}(n) & =\left[L(\phi)+\omega_m K_0\right]\rho(n)\nonumber\\
 & +\omega_m K_{+}\rho(n\mp 1)+\omega_m K_{-}\rho(n+1)\ .\label{eq:weak_measurement_kernel}
\end{align}
Note that for the $K_+$-term, we have included both of the above scenarios of either being able to distinguish the current direction or not. The resulting information processing protocols are the following. If the current direction cannot be distinguished, then $K_{+}\rho(n\mp 1)\rightarrow K_{+}\rho(n+ 1)$, such that the register $n$ is simply increased by $+1$, when having measured a nonzero current (events described by $K_\mp$). The absence of a transmitted signal corresponds to a measurement of zero current, resulting in no change in the detector count. If the current direction can be distinguished, we may engage in a different protocol, $K_{+}\rho(n\mp 1)\rightarrow K_{+}\rho(n- 1)$. Here, for a measurement of the eigenvalue $I_{-}$, the detector count goes down by one,
and and for a measurement of $I_{+}$, the detector count goes up by one.

To proceed, let us define the Fourier transform as $\rho(\xi)=\sum_n e^{i n\xi}\rho(n)$ with a new counting field $\xi$, which is distinct from, but (as we show now) to some extent related to, the original counting field $\chi$.
The Fourier transformed master equation becomes
\begin{align}
\dot{\rho}(\xi) & =\left[L(\phi)+\omega_m K_0+\omega_m K_{+}e^{i\xi}+\omega_m K_{-}e^{\mp i\xi}\right]\rho(\xi)\ .\label{eq:weak_measurement_fourier_kernel}
\end{align}
Crucially, it can be shown that for $\xi=\pi/2$, the weak measurement part in Eq.~\eqref{eq:weak_measurement_fourier_kernel} can be brought into a very similar form as the kernel in Eq.~\eqref{eq:complete_kernel_small_chi}, provided that our detector is able to distinguish between positive and negative currents, $e^{\pm i\xi}\rightarrow e^{-i\xi}$ for $K_-$. Note that setting $\xi$ to a strong nonzero value (i.e., $\xi=\pi/2$) is no problem whatsoever: the detector is here physically realized, and the experimenter will be able to directly access the classical probability distribution of measurements along the space $n$ (the space conjugate to $\xi$). Hence, the choice $\xi=\pi/2$ simply corresponds to a particular way of evaluating (post-processing) the classical information. At any rate, plugging in $\xi=\pi/2$, we then find
\begin{align}
\dot{\rho}(\pi/2) & =\left[L(\phi)+\omega_m K_0-i\frac{\omega_m \delta r}{e E_{JR}} \{I,\cdot\}\right]\rho(\pi/2)\ .\label{eq:weak_measurement_fourier_kernel_at_pihalf}
\end{align}
Indeed with the replacement
\begin{equation}
\omega_{m}\delta r\rightarrow-\chi\frac{E_{JR}}{2}\label{eq:replacement}
\end{equation}
the kernel including the weak measurement can be mapped (up to the extra term $K_0$, which we discuss in a moment) onto the kernel with small but finite $\chi$. Of course, we can therefore not hope to probe the global properties of the kernel for all $\chi$. However, this form of weak measurement can be used to ``simulate'' the presence of a small but finite counting field. Importantly, here the simulated $\chi$ enters as a system parameter influencing the dynamics, and is no longer related to the transport measurement (as the latter is encoded in the new classical counting field $\xi$). Hence, we are no longer required to expand in $\chi$, and can thus avoid any problems related to analytic continuation, and the topological phase transition shown in Fig.~\ref{fig:braids_phi}d can now be observed.

Let us now comment on the effect of aforementioned the extra term $K_0$. While the presence of $K_0$ does distort the spectrum, we observe that the braid group of the two
eigenvalues $\lambda_{\pm}$ is preserved, when setting the weak measurement parameters ($\omega_m$ and $\delta r$) to values that correspond to the value of $\chi$ [according to Eq.\ref{eq:replacement}] in Fig.~\ref{fig:braids_phi}d. As a proof of principle, we show the new eigenvalues for the weak measurement in comparison with
old ones, for finite $\chi$, in Fig.~\ref{fig:analytic_cont}b. Let us repeat that this recreation of the braid phase transition via weak measurement is only possible if the detector is able to distinguish the sign of the current (see above discussion), as otherwise, the weak measurement kernel cannot be mapped onto $L_\text{cut-off}(\phi,\chi)$. 
 
To conclude, let us provide a short interpretation of the above concept. In the absence of the weak current measurement (or transport measurement in general, $\chi=0$), the complex open system spectrum is trivial, and the eigenvalues belonging to the coherent time evolution, $\lambda_\pm$ are gapped, see Fig.~\ref{fig:open_FJE}e. This corresponds to the trivial Josephson effect. The reason why they are gapped is simply because it is in general impossible to tune the system parameters to the perfectly symmetric values $E_{JL}= E_{JR}$ and $\epsilon= 0$ (due to $U=0$). The weak measurement can close the gap, and thus correct for the ``failure'' to tune to perfect symmetry. We note that we have to set the new counting field $\xi$ to a precise value ($\pi/2$) to accomplish this. This is however no real limitation: the weak measurement has provided us with an entire array of classical information [encoded in the register $n$, $\rho(n)$, see Eq.~\eqref{eq:weak_measurement_kernel}], and setting $\xi=\pi/2$ is just a particularly chosen way to evaluate (post-process) this information. Figuratively speaking, this particular choice of post-processing the classical information ``filters out'' transport processes with integer Cooper pair charge $2e$ in favour of processes with fractional charge $e$. 

\section{Conclusions}\label{sec_conclusions}

We studied the topology of the transport properties of a generic quantum system where supercurrents and dissipative currents coincide, in terms of the transport degrees of freedom of the counting field $\chi$ and the superconducting phase bias $\phi$. We found that fractional charges defined in the full-counting statistics are related to fractional charges visible in the Josephson effect in a generic open quantum system via exceptional points defined in the 2D base space ($\chi,\phi$) -- as a matter of fact, the exceptional points can be considered as the generators of these fractional charges. While thus related, the two notions of fractional charges are nonetheless distinct in the sense that they are defined along two independent spaces. By means of the concrete model of a hetero-structure circuit, where a quantum dot is coupled to superconducting and normal metal contacts, we showed that the intricate interplay between transport measurement and nonequilibrium drive gives rise to a plethora of topological phases, surprisingly including a phase which can be interpreted as an open system version of a fractional Josephson effect, in spite of the system being composed of trivial materials.

In addition, we elucidated different flavours of full-counting statistics based on different implementations of the transport detector, and their relevance for observing different topological phases. For a continuously entangling detector, a novel type topological phase emerges, which can be interpreted as a statistical mix of fractional and integer charges defined in the counting field $\chi$. However, as we pointed out, such detectors destroy the information about the supercurrent, and thus cannot directly measure a fractional Josephson effect. A complementary approach for obtaining the full-counting statistics involves time-local measurements of the current, leaving the system be in between measurements. While this approach preserves supercurrents, it cannot detect topological transitions away from zero counting field. This prompted us to develop a third notion of full-counting statistics relying on a continuous weak measurement of the supercurrent. We sketched a proof of principle for an all-circuit implementation of such a weak measurement by means of SQUID detectors. This approach preserves supercurrents, and importantly enables us to reach topological phases at finite counting fields.

As a final note, we believe that the ``revival'' of a fractional Josephson effect by means of a weak supercurrent measurement might be an interesting effect also for actual Majorana junctions, since the gap closing in the Josephson relation may not be fully protected due to finite size effects. The applicability of weak measurement and nonequilibrium driving to induce topological protection will likely be subject of future research.

\begin{acknowledgments}
We acknowledge interesting and fruitful discussions with Maarten R.
Wegewijs and Fabian Hassler. This work has been funded by the German
Federal Ministry of Education and Research within the funding program
Photonic Research Germany under the contract number 13N14891.
\end{acknowledgments}

\appendix
\begin{widetext}

\section{Eigenmodes of system dynamics and their interpretation\label{app:eigenmodes}}

In order to find the eigenoperators belonging to $L=-i\left[H,\cdot\right]+W_N\cdot$ [as defined in Eq.~\eqref{eq:kernel_qu_sc_metal}],
we express the superoperator in terms of the operator eigenbasis of
$-i\left[H,\cdot\right]$. Due to $\Gamma_{N}\ll\left|\epsilon_{+}-\epsilon_{-}\right|$
the offdiagonal sector, given by the operator subbasis $\left|\pm\right\rangle \left\langle \mp\right|$
decouples from the diagonal subbasis $\left|+\right\rangle \left\langle +\right|,\left|-\right\rangle \left\langle -\right|,\left|1_{\sigma}\right\rangle \left\langle 1_{\sigma}\right|$,
such that the former remain eigenoperators including $W$. However,
they now acquire the additional real part $-\Gamma_{N}$ in the eigenvalue,
due to
\begin{equation}
\left\langle \pm\right|\left(W_N\left|\pm\right\rangle \left\langle \mp\right|\right)\left|\mp\right\rangle =-\Gamma_{N}.
\end{equation}
The remaining diagonal subblock $\left|+\right\rangle \left\langle +\right|,\left|-\right\rangle \left\langle -\right|,\left|1_{\sigma}\right\rangle \left\langle 1_{\sigma}\right|$
can be diagonalized separately. We find
\begin{equation}
\left\langle \pm\right|\left(W_N\left|\pm\right\rangle \left\langle \pm\right|\right)\left|\pm\right\rangle =-2\Gamma_{N}\frac{1\pm\delta}{2},
\end{equation}
and
\begin{equation}
\left\langle \pm\right|\left(W_N\left|1_{\sigma}\right\rangle \left\langle 1_{\sigma}\right|\right)\left|\pm\right\rangle =\Gamma_{N}\frac{1\mp\delta}{2},
\end{equation}
as well as
\begin{equation}
\left\langle 1_{\sigma}\right|\left(W_N\left|\pm\right\rangle \left\langle \pm\right|\right)\left|1_{\sigma}\right\rangle =\Gamma_{N}\frac{1\pm\delta}{2}\ ,
\end{equation}
with $\delta$ defined as in Eq.~\eqref{eq:detuning}.
The remaining (decoupled) diagonal subblock reads $\left[\rho\right]_{\text{diagonal}}=P_{+}\left|+\right\rangle \left\langle +\right|+\sum_{\sigma}P_{\sigma}\left|1_{\sigma}\right\rangle \left\langle 1_{\sigma}\right|+P_{-}\left|-\right\rangle \left\langle -\right|$.
For concreteness, let us assume spin degeneracy for the odd state,
$\sigma=\uparrow,\downarrow$. Writing the kernel $W$ as a matrix,
acting on the vector $\left(P_{+},P_{\uparrow},P_{\downarrow},P_{-}\right)^{T}$,
we find
\begin{equation}
\left[W\right]_{\text{diagonal}}=\left(\begin{array}{cccc}
-2\Gamma_{N}\frac{1+\delta}{2} & \Gamma_{N}\frac{1-\delta}{2} & \Gamma_{N}\frac{1-\delta}{2} & 0\\
\Gamma_{N}\frac{1+\delta}{2} & -\Gamma_{N} & 0 & \Gamma_{N}\frac{1-\delta}{2}\\
\Gamma_{N}\frac{1+\delta}{2} & 0 & -\Gamma_{N} & \Gamma_{N}\frac{1-\delta}{2}\\
0 & \Gamma_{N}\frac{1+\delta}{2} & \Gamma_{N}\frac{1+\delta}{2} & -2\Gamma_{N}\frac{1-\delta}{2}
\end{array}\right).
\end{equation}
This matrix has left and right eigenvectors, denoted with $\left|v\right)$
and $\left(w\right|$. The right eigenvectors correspond to operators
$\left|v\right)=v_{+}\left|+\right\rangle \left\langle +\right|+\sum_{\sigma}v_{\sigma}\left|1_{\sigma}\right\rangle \left\langle 1_{\sigma}\right|+v_{-}\left|-\right\rangle \left\langle -\right|$,
whereas the left eigenvectors correspond to maps from operators to
scalars, $\left(w\right|\cdot=w_{+}\left\langle +\right|\cdot\left|+\right\rangle +\sum_{\sigma}w_{\sigma}\left\langle 1_{\sigma}\right|\cdot\left|1_{\sigma}\right\rangle +w_{-}\left\langle -\right|\cdot\left|-\right\rangle $.
We here denote them in short vector notation, $\left|v\right)=\left(v_{+},v_{\uparrow},v_{\downarrow},v_{-}\right)^{T}$
and $\left(w\right|=\left(w_{+},w_{\uparrow},w_{\downarrow},w_{-}\right)$.
The left and right eigenvectors to eigenvalue $0$ are
\begin{equation}
\left(0\right|=\left(1,1,1,1\right)
\end{equation}
and
\begin{equation}
\left|0\right)=\frac{1}{4}\left(\left(1-\delta\right)^{2},1-\delta^{2},1-\delta^{2},\left(1+\delta\right)^{2}\right)^{T}.
\end{equation}
The right eigenvector corresponds to the stationary state, $\rho^{\text{st}}=\left|0\right)$,
also given in the main text. We find furthermore the eigenvalue $-2\Gamma_{N}$
with the left and right eigenvectors
\begin{align}
\left(\text{parity}\right| & =\left(1,-1,-1,1\right)-\delta^{2}\left(1,1,1,1\right)\nonumber\\
\left|\text{parity}\right) & =\frac{1}{4}\left(1,-1,-1,1\right)^{T},
\end{align}
which can be interpreted as the eigenmode connected to fermion parity
decay, along the lines given in Ref. \cite{Saptsov_2012}. Namely,
we find that the vectors can be constructed from the parity operator
$p=\left|0\right\rangle \left\langle 0\right|-\sum_{\sigma}\left|1_{\sigma}\right\rangle \left\langle 1_{\sigma}\right|+\left|2\right\rangle \left\langle 2\right|$,
which thus returns $1$ for even parity states, and $-1$ for odd
parity. In particular, the left eigenvector can be expressed as
\begin{equation}
\left(\text{parity}\right|=\left(p\right|-\left\langle p\right\rangle \left(0\right|,
\end{equation}
where $\left(p\right|=\left(1,-1,-1,1\right)$, and $\left\langle p\right\rangle =\left.\left(p\right|0\right)=\text{tr}\left[p\right]$.
Finally, there is the doubly degenerate eigenvalue $-\Gamma_{N}$.
One of them can be associated to the spin decay, as its left and right
eigenvectors are
\begin{align}
\left(\text{spin}\right| & =\left(0,1,-1,0\right)\nonumber\\
\left|\text{spin}\right) & =\frac{1}{2}\left(0,1,-1,0\right)^{T},
\end{align}
which can obviously related to the spin operator $s=\sum_{\sigma}\sigma\left|1_{\sigma}\right\rangle \left\langle 1_{\sigma}\right|$.
The remaining set of left and right eigenvectors can be related to
what we referred to in the main text as the pseudocharge,
\begin{align}
\left(\text{pseudocharge}\right| & =\left(1,0,0,-1\right)+\delta\left(1,1,1,1\right)\nonumber\\
\left|\text{pseudocharge}\right) & =\frac{1}{2}\left(1,0,0,-1\right)^{T},
\end{align}
related to the operator $\widehat{c}=\left|+\right\rangle \left\langle +\right|-\left|-\right\rangle \left\langle -\right|$. Likewise,
we may express the left eigenvector as
\begin{equation}
\left(\text{pseudocharge}\right|=\left(c\right|-\left\langle \widehat{c}\right\rangle \left(0\right|,
\end{equation}
with $\left(c\right|=\left(1,0,0,-1\right)$.

\section{Exceptional points for analysis of topological transitions\label{app:excep_points}}

In this appendix, we derive the expressions for the exceptional points
mentioned in the main text, which enable a quick numerical analysis
of the topological maps in Figs. \ref{fig:map_chi}. For this purpose,
we need the eigenvalues of $L\left(\chi,\phi\right)$, defined in
Eq.~(\ref{eq:L_chi}), which can be computed analytically. Defining
the operator basis $\left|0\right\rangle \left\langle 0\right|,\left|0\right\rangle \left\langle 2\right|,\left|2\right\rangle \left\langle 0\right|,\left|2\right\rangle \left\langle 2\right|,\left|\uparrow\right\rangle \left\langle \uparrow\right|,\left|\downarrow\right\rangle \left\langle \downarrow\right|$,
the superoperator $L\left(\chi,\phi\right)$ can be written in form
of the matrix
\begin{align}
L\left(\chi,\phi\right) & =\left(\begin{array}{cccccc}
-2\Gamma_{N} & -i\alpha_{1} & i\alpha_{2} & 0 & 0 & 0\\
-i\alpha_{4} & i\left(2\epsilon+U\right)-\Gamma_{N} & 0 & i\alpha_{2} & 0 & 0\\
i\alpha_{3} & 0 & -i\left(2\epsilon+U\right)-\Gamma_{N} & -i\alpha_{1} & 0 & 0\\
0 & i\alpha_{3} & -i\alpha_{4} & 0 & \Gamma_{N} & \Gamma_{N}\\
\Gamma_{N} & 0 & 0 & 0 & -\Gamma_{N} & 0\\
\Gamma_{N} & 0 & 0 & 0 & 0 & -\Gamma_{N}
\end{array}\right),
\end{align}
with
\begin{align}
\alpha_{1} & =\frac{1}{2}\left(E_{JL}+E_{JR}e^{i\phi}e^{i\chi}\right)\nonumber\\
\alpha_{2} & =\frac{1}{2}\left(E_{JL}+E_{JR}e^{-i\phi}e^{i\chi}\right)\nonumber\\
\alpha_{3} & =\frac{1}{2}\left(E_{JL}+E_{JR}e^{i\phi}e^{-i\chi}\right)\nonumber\\
\alpha_{4} & =\frac{1}{2}\left(E_{JL}+E_{JR}e^{-i\phi}e^{-i\chi}\right).
\end{align}
Of course for real counting fields $\chi$ and phases $\phi$, $\alpha_{4}=\alpha_{1}^{*}$
and $\alpha_{3}=\alpha_{2}^{*}$. However, in a moment, we will extend
the system to complex $\chi$ and $\phi$, respectively, where these
relationships are no longer valid.

For the above matrix, we find the doubly degenerate eigenvalues $-\Gamma_{N}$,
associated to the decay of spin and pseudocharge (see main text, and
Appendix \ref{app:eigenmodes}), which therefore depend neither on
$\chi$, nor on $\phi$. These eigenvalues can therefore not partake
in any braid phase transitions. The remaining four eigenvalues are
\begin{equation}
-\Gamma_{N}\pm\frac{1}{\sqrt{2}}\sqrt{A-\sqrt{B}}\quad-\Gamma_{N}\pm\frac{1}{\sqrt{2}}\sqrt{A+\sqrt{B}}
\end{equation}
with
\begin{align}
A & =\Gamma_{N}^{2}-\epsilon^{2}-2\alpha_{1}\alpha_{4}-2\alpha_{2}\alpha_{3}\\
B & =16\alpha_{1}\alpha_{2}\left(\alpha_{3}\alpha_{4}+\Gamma_{N}^{2}\right)-4\alpha_{1}\alpha_{4}\left(\Gamma_{N}+i\epsilon\right)^{2}\\
 & -4\alpha_{2}\alpha_{3}\left(\Gamma_{N}-i\epsilon\right)^{2}+\left(\Gamma_{N}^{2}+\epsilon^{2}\right)^{2}.
\end{align}
The exceptional points (degeneracies) can be found through the equations
\begin{equation}
B=0\quad\text{and}\quad A^{2}-B=0.
\end{equation}
In order to find the exceptional points for transitions in $\chi$
or in $\phi$, we either have to make the replacement $e^{\pm i\chi}\rightarrow z^{\pm1}$
or $e^{\pm i\phi}\rightarrow\widetilde{z}^{\pm1}$ (as indicated in
the main text). Consequently, both conditions lead to polynomial equations
of fourth order, via
\begin{align}
z^{2}B\left(z\right) & =p_{4}z^{4}+p_{3}z^{3}+p_{2}z^{2}+p_{1}z+p_{0}\nonumber\\
z^{2}\left[A^{2}\left(z\right)-B\left(z\right)\right] & =q_{4}z^{4}+q_{3}z^{3}+q_{2}z^{2}+q_{1}z+q_{0}\ .
\end{align}
We thus recover Eq.~\eqref{eq_EP_quartic} in the main text, and its variant for $\widetilde{z}$, where we have different coefficients
$\widetilde{p}_{k}$ and $\widetilde{q}_{k}$. Note that these are two independent equations, that is, we receive a set of roots for $z$ ($\widetilde{z}$) for the quartic equation with $p_i$ ($\widetilde{p}_i$) and another set with $q_i$ ($\widetilde{q}_i$). We find the coefficients
\begin{align}
p_{0} & =E_{JL}^{2}E_{JR}^{2}\nonumber\\
p_{1} & =2E_{JL}E_{JR}\left[\left(E_{JL}^{2}+E_{JR}^{2}+\epsilon^{2}-\Gamma_{N}^{2}\right)\cos\left(\phi\right)-2\Gamma_{N}\epsilon\sin\left(\phi\right)\right]\nonumber\\
p_{2} & =\left[E_{JL}^{2}+E_{JR}^{2}+\Gamma_{N}^{2}+\epsilon^{2}\right]^{2}-4E_{JR}^{2}\Gamma_{N}^{2}+2E_{JL}^{2}E_{JR}^{2}\cos\left(2\phi\right)\nonumber\\
p_{3} & =2E_{JL}E_{JR}\left[\left(E_{JL}^{2}+E_{JR}^{2}+\epsilon^{2}+3\Gamma_{N}^{2}\right)\cos\left(\phi\right)+2\Gamma_{N}\epsilon\sin\left(\phi\right)\right]\nonumber\\
p_{4} & =E_{JR}^{2}\left(E_{JL}^{2}+4\Gamma_{N}^{2}\right)
\end{align}
and
\begin{align}
q_{0} & =-E_{JL}^{2}E_{JR}^{2}\sin^{2}\left(\phi\right)\nonumber\\
q_{1} & =4E_{JL}E_{JR}\Gamma_{N}\epsilon\sin\left(\phi\right)\nonumber\\
q_{2} & =E_{JL}^{2}\left(E_{JR}^{2}-4\Gamma_{N}^{2}\right)-4\Gamma_{N}^{2}\epsilon^{2}-E_{JL}^{2}E_{JR}^{2}\cos\left(2\phi\right)\nonumber\\
q_{3} & =-4E_{JL}E_{JR}\Gamma_{N}\left[2\Gamma_{N}\cos\left(\phi\right)+\epsilon\sin\left(\phi\right)\right]\nonumber\\
q_{4} & =-E_{JR}^{2}\left[E_{JL}^{2}\sin^{2}\left(\phi\right)+4\Gamma_{N}^{2}\right].
\end{align}
For the set of equations with $\widetilde{z}$, we find the coefficients,
\begin{align}
\widetilde{p}_{0} & =E_{JL}^{2}E_{JR}^{2}\nonumber\\
\widetilde{p}_{1} & =2E_{JL}E_{JR}\left[\left(E_{JL}^{2}+E_{JR}^{2}+\Gamma_{N}^{2}+\epsilon^{2}\right)\cos\left(\chi\right)+2\Gamma_{N}\left(i\Gamma_{N}-\epsilon\right)\sin\left(\chi\right)\right]\nonumber\\
\widetilde{p}_{2} & =\left[E_{JL}^{2}+E_{JR}^{2}+\Gamma_{N}^{2}+\epsilon^{2}\right]^{2}-4E_{JR}^{2}\Gamma_{N}^{2}\left(1-e^{i\chi}\right)+2E_{JL}^{2}E_{JR}^{2}\cos\left(2\phi\right)\nonumber\\
\widetilde{p}_{3} & =2E_{JL}E_{JR}\left[\left(E_{JL}^{2}+E_{JR}^{2}+\Gamma_{N}^{2}+\epsilon^{2}\right)\cos\left(\chi\right)+2\Gamma_{N}\left(i\Gamma_{N}+\epsilon\right)\sin\left(\chi\right)\right]\nonumber\\
\widetilde{p}_{4} & =E_{JL}^{2}E_{JR}^{2}
\end{align}
and
\begin{align}
\widetilde{q}_{0} & =-E_{JL}^{2}E_{JR}^{2}\sin^{2}\left(\chi\right)\nonumber\\
\widetilde{q}_{1} & =-4E_{JL}E_{JR}\Gamma_{N}\left[\Gamma_{N}e^{i\chi}-\epsilon\sin\left(\chi\right)\right]\nonumber\\
\widetilde{q}_{2} & =E_{JL}^{2}\left(E_{JR}^{2}-4\Gamma_{N}^{2}\right)-4\Gamma_{N}^{2}\left(\epsilon^{2}+e^{i2\chi}E_{JR}^{2}\right)-E_{JL}^{2}E_{JR}^{2}\cos\left(2\chi\right)\nonumber\\
\widetilde{q}_{3} & =-4E_{JL}E_{JR}\Gamma_{N}\left[\Gamma_{N}e^{i\chi}+\epsilon\sin\left(\chi\right)\right]\nonumber\\
\widetilde{q}_{4} & =-E_{JL}^{2}E_{JR}^{2}\sin^{2}\left(\chi\right).
\end{align}
All that is left, is using the quartic formula to find the positions of the exceptional points in $z$ or $\widetilde{z}$.

\section{Extracting derivatives of eigenvalues from experiments\label{app:analytic_continuation}}

In the main text, we argue that the $\chi$-derivatives of eigenvalues $\lambda$ can be extracted from experimental data. We here briefly explain this statement. It is rooted in the first definition of FCS with time-local current measurements, as explained in Sec.~\ref{sec:time-resolved_FCS} of the main text. Here,
one measures the cumulants, as defined in Eqs.~\eqref{eq_current_C1} and~\eqref{eq_noise_C2}. The moment and cumulant generating functions
can be expressed in terms of the eigenmodes of $L(\chi,\phi)$ as
\begin{align}
m\left(\chi,\tau\right) & =e^{\tau c\left(\chi,\phi,\tau\right)}\equiv\text{tr}\left[e^{L\left(\chi,\phi\right)\tau}\rho_{0}\right]\nonumber\\
 & =\sum_{n}e^{\lambda_{n}\left(\chi,\phi\right)\tau}\underbrace{\text{tr}\left[\left|n\left(\chi,\phi\right)\right)\left(n\left(\chi,\phi\right)\right|\rho_{0}\right]}_{e^{\alpha_{n}\left(\chi,\phi\right)}}\nonumber\\
 & =\sum_{n}e^{\lambda_{n}\left(\chi,\phi\right)\tau+\alpha_{n}\left(\chi,\phi\right)}\nonumber\\
 & =e^{\lambda_{0}\left(\chi,\phi\right)\tau+\alpha_{0}\left(\chi,\phi\right)}+\sum_{n\neq0}e^{\lambda_{n}\left(\chi,\phi\right)\tau+\alpha_{n}\left(\chi,\phi\right)}
\end{align}
where $\lambda_{n}(\chi,\phi)$ are the eigenvalues of $L(\chi,\phi)$
while $\left|n\left(\chi,\phi\right)\right)$ and $\left(n\left(\chi,\phi\right)\right|$
are its right and left eigenvectors, respectively. We know that $\lambda_{0}\left(0,\phi\right)=0$
and $\alpha_{0}\left(0,\phi\right)=0$ if $\rho_{0}$ is the stationary
state. The moment and cumulant generating functions can be expanded
about $\chi=0$. For notational simplicity, we omit the addition of the $(-i)^k$ prefactor for the $k$-th cumulant, and we likewise neglect the elementary charge prefactor $e$. Note that the physically measurable cumulants $C_k$ in the main text and the below defined $c_k$ are related as $C_k=(-ie)^k c_k$. At any rate, we find
\begin{align}
m\left(\chi,\phi,\tau\right) & \approx m_{0}\left(\phi,\tau\right)+\chi m_{1}\left(\phi,\tau\right)+\frac{1}{2}\chi^{2}m_{2}\left(\phi,\tau\right)+\ldots\nonumber\\
 & m_{k}\left(\phi,\tau\right)=\left.\partial_{\chi}^{k}m\left(\chi,\phi,\tau\right)\right|_{\chi\rightarrow0}\nonumber\\
c\left(\chi,\phi,\tau\right) & \approx c_{0}\left(\phi,\tau\right)+\chi c_{1}\left(\phi,\tau\right)+\frac{1}{2}\chi^{2}c_{2}\left(\phi,\tau\right)+\ldots\nonumber\\
 & c_{k}\left(\phi,\tau\right)=\left.\partial_{\chi}^{k}c\left(\chi,\phi,\tau\right)\right|_{\chi\rightarrow0}
\end{align}
Now we can define the derivatives of the cumulant generating function
in terms of the derivatives of the moment generating function via the natural logarithm
\begin{align}
\frac{1}{\tau}\ln\left[1+\chi m_{1}\left(\phi,\tau\right)+\frac{1}{2}\chi^{2}m_{2}\left(\phi,\tau\right)+\ldots\right] & =c_{0}\left(\phi,\tau\right)+\chi c_{1}\left(\phi,\tau\right)+\frac{1}{2}\chi^{2}c_{2}\left(\phi,\tau\right)+\ldots\nonumber\\
\ln\left[1+\chi m_{1}\left(\phi,\tau\right)+\frac{1}{2}\chi^{2}m_{2}\left(\phi,\tau\right)+\ldots\right] & \approx\chi m_{1}\left(\phi,\tau\right)+\frac{1}{2}\chi^{2}\left[m_{2}\left(\phi,\tau\right)-m_{1}^{2}\left(\phi,\tau\right)\right]+\ldots \ ,
\end{align}
where we have used the Maclaurin series expansion $\ln(1+x)=x-x^{2}/2+x^{3}/3-...$. 
An order by order comparison yields the well-known relationships
\begin{align}
c_{0}\left(\phi,\tau\right) & =0\nonumber\\
c_{1}\left(\phi,\tau\right) & =\frac{1}{\tau}m_{1}\left(\phi,\tau\right)\nonumber\\
c_{2}\left(\phi,\tau\right) & =\frac{1}{\tau}\left[m_{2}\left(\phi,\tau\right)-m_{1}^{2}\left(\phi,\tau\right)\right]
\end{align}
To proceed we now have to find the relation between the derivatives of moment generating
function and the derivatives of eigenvalues. For the latter, we get
\begin{align}
\lambda_{n}\left(\chi,\phi\right)\tau+\alpha_{n}\left(\chi,\phi\right) & \approx\left[\lambda_{n}^{\left(0\right)}\left(\phi\right)+\chi\lambda_{n}^{\left(1\right)}\left(\phi\right)+\frac{1}{2}\chi^{2}\lambda_{n}^{\left(2\right)}\left(\phi\right)\right]\tau+\left[\alpha_{n}^{\left(0\right)}\left(\phi\right)+\chi\alpha_{n}^{\left(1\right)}\left(\phi\right)+\frac{1}{2}\chi^{2}\alpha_{n}^{\left(2\right)}\left(\phi\right)\right]+\ldots
\end{align}
where $\lambda_{n}^{(k)}(\phi)=\left.\partial_{\chi}\lambda_{n}(\chi,\phi)\right|_{\chi\to0}$, and the
same for $\alpha_{n}^{(k)}(\phi)$. Plugging this expansion of $\lambda$ and $\alpha$ into the definition of the moment generating function,
we get
\begin{align}
m_0(\phi,\tau) &=1\\
m_{1}(\phi,\tau) & =\sum_{n}e^{\lambda_{n}^{\left(0\right)}\left(\phi\right)\tau+\alpha_{n}^{\left(0\right)}\left(\phi\right)}\left[\lambda_{n}^{\left(1\right)}\left(\phi\right)\tau+\alpha_{n}^{\left(1\right)}\left(\phi\right)\right]\nonumber\\
m_{2}(\phi,\tau) & =\sum_{n}e^{\lambda_{n}^{\left(0\right)}\left(\phi\right)\tau+\alpha_{n}^{\left(0\right)}\left(\phi\right)}\left[\lambda_{n}^{\left(2\right)}\left(\phi\right)\tau+\alpha_{n}^{\left(2\right)}\left(\phi\right)+\left(\lambda_{n}^{\left(1\right)}\left(\phi\right)\tau+\alpha_{n}^{\left(1\right)}\left(\phi\right)\right)^{2}\right]\ .
\end{align}
Finally we can write the cumulants in terms of derivatives of eigenvalues
\begin{align}
c_{1}\left(\phi,\tau\right)
 & =\sum_{n}e^{\lambda_{n}^{\left(0\right)}\left(\phi\right)\tau}e^{\alpha_{n}^{\left(0\right)}\left(\phi\right)}\left[\lambda_{n}^{\left(1\right)}\left(\phi\right)+\frac{1}{\tau}\alpha_{n}^{\left(1\right)}\left(\phi\right)\right]\nonumber\\
c_{2}\left(\phi,\tau\right) & =\sum_{n}e^{\lambda_{n}^{\left(0\right)}\left(\phi\right)\tau}e^{\alpha_{n}^{\left(0\right)}\left(\phi\right)}\left[\lambda_{n}^{\left(2\right)}\left(\phi\right)+2\lambda_{n}^{\left(1\right)}\left(\phi\right)\alpha_{n}^{\left(1\right)}\left(\phi\right)+\left[\lambda_{n}^{\left(1\right)}\left(\phi\right)\right]^{2}\tau+\frac{1}{\tau}\left(\alpha_{n}^{\left(2\right)}\left(\phi\right)+\left[\alpha_{n}^{\left(1\right)}\left(\phi\right)\right]^{2}\right)\right]\nonumber\\
 & -\sum_{nn'}e^{\lambda_{n}^{\left(0\right)}\left(\phi\right)\tau}e^{\lambda_{n'}^{\left(0\right)}\left(\phi\right)\tau}\Biggl[e^{\alpha_{n}^{\left(0\right)}\left(\phi\right)}\lambda_{n}^{\left(1\right)}\left(\phi\right)e^{\alpha_{n'}^{\left(0\right)}\left(\phi\right)}\lambda_{n'}^{\left(1\right)}\left(\phi\right)\tau+e^{\alpha_{n}^{\left(0\right)}\left(\phi\right)}\lambda_{n}^{\left(1\right)}\left(\phi\right)e^{\alpha_{n'}^{\left(0\right)}\left(\phi\right)}\alpha_{n'}^{\left(1\right)}\left(\phi\right)\nonumber\\
 & +e^{\alpha_{n}^{\left(0\right)}\left(\phi\right)}\alpha_{n}^{\left(1\right)}\left(\phi\right)e^{\alpha_{n'}^{\left(0\right)}\left(\phi\right)}\lambda_{n'}^{\left(1\right)}\left(\phi\right)+\frac{1}{\tau}e^{\alpha_{n}^{\left(0\right)}\left(\phi\right)}\alpha_{n}^{\left(1\right)}\left(\phi\right)e^{\alpha_{n'}^{\left(0\right)}\left(\phi\right)}\alpha_{n'}^{\left(1\right)}\left(\phi\right)\Biggr]
\end{align}
Just to briefly confirm,
in the long time limit $\tau\rightarrow\infty$,
the second cumulant becomes $c_{2}(\phi,\tau)=-\lambda_{0}^{(2)}(\phi)$.
Now, the idea is the following. The quantities $c_1$ and $c_2$ are measurable as a function of $\tau$, when performing a finite frequency evaluation of current and noise. Then, all quantities which have a distinct time-evolution (be it due to a different exponential decay due to $\lambda_n$ or due to a prefactor with a different power-law in $\tau$) can be distinguished and extracted in principle by fitting of the $\tau$-dependent data, by an appropriately chosen fitting function.

Consequently, in addition to the decay rates $\lambda_n^{0}$, the following terms can be individually extracted from the experimental data.
For the first cumulant, these are
\begin{align}
    o_{1,n} =& e^{\alpha_{n}^{\left(0\right)}\left(\phi\right)}\lambda_{n}^{\left(1\right)}\left(\phi\right)\\
    o_{2,n} =& e^{\alpha_{n}^{\left(0\right)}\left(\phi\right)}\alpha_{n}^{\left(1\right)}\left(\phi\right) \ .
\end{align}
From the second cumulant, we can independently extract
\begin{align}
    o_{3,n} =& e^{\alpha_{n}^{\left(0\right)}\left(\phi\right)}\lambda_{n}^{\left(2\right)}\left(\phi\right)+2e^{\alpha_{n}^{\left(0\right)}\left(\phi\right)}\lambda_{n}^{\left(1\right)}\left(\phi\right)\alpha_{n}^{\left(1\right)}\left(\phi\right)\\  o_{4,n} =& e^{\alpha_{n}^{\left(0\right)}\left(\phi\right)}\left[\lambda_{n}^{\left(1\right)}\left(\phi\right)\right]^{2}\\                                         o_{5,n} =& e^{\alpha_{n}^{\left(0\right)}\left(\phi\right)}\left(\alpha_{n}^{\left(2\right)}\left(\phi\right)+\left[\alpha_{n}^{\left(1\right)}\left(\phi\right)\right]^{2}\right).
\end{align}
Finally by taking different combination of these expressions we can
get the first and second order corrections of \textit{all} eigenvalues, that is
\begin{equation}
    \lambda_{n}^{(1)}(\phi)=\frac{o_{4,n}}{o_{1,n}}\ ,
\end{equation}
and
\begin{equation}
    \lambda_{n}^{(2)}(\phi)=\frac{\left(o_{1,n}o_{3,n}-2o_{2,n}o_{4,n}\right) o_{4,n}}{o_{n,1}^3}\ .
\end{equation} 
These individual results can be stitched together to get $\lambda_n(\chi)\approx \lambda_n^{(0)}+\chi \lambda_n^{(1)}+\frac{1}{2}\chi^2 \lambda_n^{(2)}$, which we plot in Fig.~\ref{fig:analytic_cont}a in the main text for $\lambda_{\pm}$. While in principle, this procedure allows us to  analytically continue the eigenvalues from $\chi=0$ to finite $\chi$, we see that this continuation fails to converge if there is a topological phase transition from zero $\chi$ to finite $\chi$, unless one measures cumulants up to infinite order in $k$, which is a prohibitive requirement.

\section{Calculation of the transmission coefficient for the squid detector\label{app:transmission_coeff}}

In the main text, we describe an all-circuit realization of continuous weak measurement of the supercurrent. The decisive figure of merit is the transmission coefficient of incoming waves towards the SQUID detector. To calculate this transmission coefficient, we start from the Hamiltonian description of the SQUID detector given in Eqs.~(\ref{eq:SQUID_hamiltonian_HL}-\ref{eq:SQUID_hamiltonian_V}), and
first diagonalize the Hamiltonians for the left and right conductor
lines, i.e. $H_{L}$ and $H_{R}$, using the following mode expansion
\begin{align}
q_{j,L/R} & =-\frac{i}{\sqrt{2}}\int_{0}^{\pi}\frac{dk}{\pi}\left(\frac{C_{0}}{L_{0}}\right)^{1/4}\sqrt{2\sin\left(\frac{\left|k\right|}{2}\right)}\cos\left(k(j-1/2)\right)\left[a_{k,L/R}-a_{k,L/R}^{\dagger}\right]\nonumber\\
\frac{\varphi_{j,L/R}}{2e} & =\frac{1}{\sqrt{2}}\int_{0}^{\pi}\frac{dk}{\pi}\left(\frac{L_{0}}{C_{0}}\right)^{1/4}\frac{1}{\sqrt{2\sin\left(\frac{\left|k\right|}{2}\right)}}\cos\left(k(j-1/2)\right)\left[a_{k,L/R}+a_{k,L/R}^{\dagger}\right]
\end{align}
The free Hamiltonians becomes
\begin{align}
H_{0} & =H_{L}\otimes\mathbb{I}_{R}+\mathbb{I}_{L}\otimes H_{R}=\omega_{0}\int_{0}^{\pi}\frac{dk}{\pi}\sin\left(\frac{k}{2}\right)a_{k,L}^{\dagger}a_{k,L}\otimes\mathbb{I}_{R}+\mathbb{I}_{L}\otimes\omega_{0}\int_{0}^{\pi}\frac{dk}{\pi}\sin\left(\frac{k}{2}\right)a_{k,R}^{\dagger}a_{k,R}\ .
\end{align}
We note that in the continuum limit (vanishing size of islands $j$), we recover a linear dispersion relation $\sim \omega_0 k$. For now we keep finite size effects, and take the continuum limit at an appropriate later time.

Consequently, the interaction term can be expressed in terms of these bosonic operators as follows,
\begin{align}
V & =\gamma\left(\frac{\varphi_{1}-\varphi_{0}}{2e}\right)^{2}\nonumber\\
 & =\frac{\gamma}{4}\sqrt{\frac{L_{0}}{C_{0}}}\int_{0}^{\pi}\int_{0}^{\pi}\frac{dkdk'}{(2\pi)^{2}}\frac{\cos\left(\frac{k}{2}\right)\cos\left(\frac{k'}{2}\right)}{\sqrt{\sin\left(\frac{k}{2}\right)\sin\left(\frac{k'}{2}\right)}}\left[\mathbb{I}_{L}\otimes a_{k,R}+\mathbb{I}_{L}\otimes a_{k,R}^{\dagger}-a_{k,L}\otimes\mathbb{I}_{R}-a_{k,L}^{\dagger}\otimes\mathbb{I}_{R}\right]\nonumber\\
 & *\left[\mathbb{I}_{L}\otimes a_{k',R}+\mathbb{I}_{L}\otimes a_{k',R}^{\dagger}-a_{k',L}\otimes\mathbb{I}_{R}-a_{k',L}^{\dagger}\otimes\mathbb{I}_{R}\right]\nonumber\\
 & =\frac{\gamma}{4}\sqrt{\frac{L_{0}}{C_{0}}}\int_{0}^{\pi}\int_{0}^{\pi}\frac{dkdk'}{(2\pi)^{2}}\frac{\cos\left(\frac{k}{2}\right)\cos\left(\frac{k'}{2}\right)}{\sqrt{\sin\left(\frac{k}{2}\right)\sin\left(\frac{k'}{2}\right)}}\biggl[\mathbb{I}_{L}\otimes\left(a_{k,R}a_{k',R}+a_{k,R}^{\dagger}a_{k',R}^{\dagger}\right)+2\mathbb{I}_{L}\otimes a_{k,R}^{\dagger}a_{k',R}-a_{k,L}\otimes a_{k',R}\nonumber\\
 & -a_{k,L}^{\dagger}\otimes a_{k',R}^{\dagger}-a_{k,L}^{\dagger}\otimes a_{k',R}-a_{k,L}\otimes a_{k',R}^{\dagger}-a_{k',L}\otimes a_{k,R}-a{}_{k',L}^{\dagger}\otimes a_{k,R}^{\dagger}-a_{k',L}\otimes a_{k,R}^{\dagger}\nonumber\\
 & -a_{k',L}^{\dagger}\otimes a_{k,R}+\left(a_{k,L}a_{k',L}+a_{k,L}^{\dagger}a_{k',L}^{\dagger}\right)\otimes\mathbb{I}_{R}+2a_{k,L}^{\dagger}a_{k',L}\otimes\mathbb{I}_{R}\biggr]
\end{align}
We can now deploy the following important simplification. We will be considering an incoming signal at a certain energy focussing on the limit of elastic interaction (neglecting a small chance that the boson may be absorbed by the main circuit). In addition, we perform a rotation wave approximation, discarding the pair-wise creation (annihilation) terms $\sim a^\dagger a^\dagger$ ($\sim a a$). This allows us to work in the single particle picture, the relevant states being
 $\left|0,k\right\rangle $ and $\left|k,0\right\rangle $,
where the first (second) states corresponds to a single particle eigenstate of $H_{L}$ ($H_{R}$). The corresponding Green's functions are
\begin{align}
\left(G_{k,k'}\right)_{LL} & =\left\langle k,0\right|\frac{1}{E-H+i0^{+}}\left|k',0\right\rangle \nonumber\\
\left(G_{k,k'}\right)_{LR} & =\left\langle k,0\right|\frac{1}{E-H+i0^{+}}\left|0,k'\right\rangle \nonumber\\
\left(G_{k,k'}\right)_{RL} & =\left\langle 0,k\right|\frac{1}{E-H+i0^{+}}\left|k',0\right\rangle \nonumber\\
\left(G_{k,k'}\right)_{RR} & =\left\langle 0,k\right|\frac{1}{E-H+i0^{+}}\left|0,k'\right\rangle 
\end{align}
According to the Fisher-Lee formula~\cite{Fisher_and_Lee_1981}, for the transmission coefficient we need to find $\left(G_{k,k'}\right)_{RL}$.
We can write the Dyson equation as
\begin{align}
\left(G_{k,k'}\right)_{RL}=\left\langle 0,k\right|\frac{1}{E-H+i0^{+}}\left|k',0\right\rangle  & =\left\langle 0,k\right|\frac{1}{E-H_{0}+i0^{+}}\left|k',0\right\rangle +\left\langle 0,k\right|\frac{1}{E-H_{0}+i0^{+}}V\frac{1}{E-H+i0^{+}}\left|k',0\right\rangle 
\label{eq:Dyson_eq}
\end{align}
Since we consider only processes in $V$ conserving the total boson number, we can insert the identity operator for single
particle subspace
\begin{align}
\mathbb{I}_{L} & =\int_{0}^{\pi}\frac{dk}{2\pi}\left|k,0\right\rangle \left\langle k,0\right|\nonumber\\
\mathbb{I} & =\mathbb{I}_{L}\oplus\mathbb{I}_{R}=\int_{0}^{\pi}\frac{dk_{1}}{2\pi}\left|k_{1},0\right\rangle \left\langle k_{1},0\right|+\int_{0}^{\pi}\frac{dk_{2}}{2\pi}\left|0,k_{2}\right\rangle \left\langle 0,k_{2}\right| \ .
\end{align}
to get
\begin{align}
\left(G_{k,k'}\right)_{RL} & =\left\langle 0,k\right|\frac{1}{E-H_{0}+i0^{+}}\left|k',0\right\rangle \nonumber\\
 & +\left\langle 0,k\right|\frac{1}{E-H_{0}+i0^{+}}\int_{0}^{\pi}\frac{dk_{1}}{2\pi}\left|k_{1},0\right\rangle \left\langle k_{1},0\right|V\int_{0}^{\pi}\frac{dk_{3}}{2\pi}\left|k_{3},0\right\rangle \left\langle k_{3},0\right|\frac{1}{E-H+i0^{+}}\left|k',0\right\rangle \nonumber\\
 & +\left\langle 0,k\right|\frac{1}{E-H_{0}+i0^{+}}\int_{0}^{\pi}\frac{dk_{1}}{2\pi}\left|k_{1},0\right\rangle \left\langle k_{1},0\right|V\int_{0}^{\pi}\frac{dk_{4}}{2\pi}\left|0,k_{4}\right\rangle \left\langle 0,k_{4}\right|\frac{1}{E-H+i0^{+}}\left|k',0\right\rangle \nonumber\\
 & +\left\langle 0,k\right|\frac{1}{E-H_{0}+i0^{+}}\int_{0}^{\pi}\frac{dk_{2}}{2\pi}\left|0,k_{2}\right\rangle \left\langle 0,k_{2}\right|V\int_{0}^{\pi}\frac{dk_{3}}{2\pi}\left|k_{3},0\right\rangle \left\langle k_{3},0\right|\frac{1}{E-H+i0^{+}}\left|k',0\right\rangle \nonumber\\
 & +\left\langle 0,k\right|\frac{1}{E-H_{0}+i0^{+}}\int_{0}^{\pi}\frac{dk_{2}}{2\pi}\left|0,k_{2}\right\rangle \left\langle 0,k_{2}\right|V\int_{0}^{\pi}\frac{dk_{4}}{2\pi}\left|0,k_{4}\right\rangle \left\langle 0,k_{4}\right|\frac{1}{E-H+i0^{+}}\left|k',0\right\rangle 
\end{align}
Assuming weak tunneling, we may deploy a perturbation theory up to first order in the interaction $V$. To this end, we set $H=H_{0}$
on the right side of Eq.\ref{eq:Dyson_eq}, to get
\begin{align}
\left(G_{k,k'}\right)_{RL}  & \approx\left\langle 0,k\right|\frac{1}{E-H_{0}+i0^{+}}\left|k',0\right\rangle +\left\langle 0,k\right|\frac{1}{E-H_{0}+i0^{+}}V\frac{1}{E-H_{0}+i0^{+}}\left|k',0\right\rangle \nonumber\\
 & =\frac{1}{E-\omega_{k}+i0^{+}}\left\langle 0,k\right|V\left|k',0\right\rangle \frac{1}{E-\omega_{k'}+i0^{+}}
\end{align}
Since without the interaction term $V$, there cannot be any coupling between the left and right side of the system, the free Green's function connecting the left and right momenta
is zero, i.e.
\begin{equation}
\left\langle 0,k\right|\frac{1}{E-H_{0}+i0^{+}}\left|k',0\right\rangle =0\ .
\end{equation}
This leaves us with the task of computing the interaction term
\begin{align}\nonumber
\left\langle 0,k\right|V\left|k',0\right\rangle  & =\frac{\gamma}{4}\sqrt{\frac{L_{0}}{C_{0}}}\int_{0}^{\pi}\int_{0}^{\pi}\frac{dk_{1}dk_{2}}{(2\pi)^{2}}\frac{\cos\left(\frac{k_{1}}{2}\right)\cos\left(\frac{k_{2}}{2}\right)}{\sqrt{\sin\left(\frac{k_{1}}{2}\right)\sin\left(\frac{k_{2}}{2}\right)}}\left\langle 0\right|\otimes\left\langle 0\right|a_{k,R}\left[\mathbb{I}_{L}\otimes a_{k_{1},R}+\mathbb{I}_{L}\otimes a_{k_{1},R}^{\dagger}\right.\\ &\left.-a_{k_{1},L}\otimes\mathbb{I}_{R}-a_{k_{1},L}^{\dagger}\otimes\mathbb{I}_{R}\right]\left[\mathbb{I}_{L}\otimes a_{k_{2},R}+\mathbb{I}_{L}\otimes a_{k_{2},R}^{\dagger}-a_{k_{2},L}\otimes\mathbb{I}_{R}-a_{k_{2},L}^{\dagger}\otimes\mathbb{I}_{R}\right]a_{k',L}^{\dagger}\left|0\right\rangle \otimes\left|0\right\rangle \nonumber\\
 & =\frac{\gamma}{4}\sqrt{\frac{L_{0}}{C_{0}}}\int_{0}^{\pi}\int_{0}^{\pi}\frac{dk_{1}dk_{2}}{(2\pi)^{2}}\frac{\cos\left(\frac{k_{1}}{2}\right)\cos\left(\frac{k_{2}}{2}\right)}{\sqrt{\sin\left(\frac{k_{1}}{2}\right)\sin\left(\frac{k_{2}}{2}\right)}}\left\langle 0\right|\otimes\left\langle 0\right|a_{k,R}\biggl[\mathbb{I}_{L}\otimes a_{k_{1},R}a_{k_{2},R}+\mathbb{I}_{L}\otimes a_{k_{1},R}a_{k_{2},R}^{\dagger}\nonumber\\
 & -a_{k_{2},L}\otimes a_{k_{1},R}-a_{k_{2},L}^{\dagger}\otimes a_{k_{1},R}+\mathbb{I}_{L}\otimes a_{k_{1},R}^{\dagger}a_{k_{2},R}+\mathbb{I}_{L}\otimes a_{k_{1},R}^{\dagger}a_{k_{2},R}^{\dagger}-a_{k_{2},L}\otimes a_{k_{1},R}^{\dagger}\nonumber\\
 & -a_{k_{2},L}^{\dagger}\otimes a_{k_{1},R}^{\dagger}-a_{k_{1},L}\otimes a_{k_{2},R}-a_{k_{1},L}\otimes a_{k_{2},R}^{\dagger}+a_{k_{1},L}a_{k_{2},L}\otimes\mathbb{I}_{R}+a_{k_{1},L}a_{k_{2},L}^{\dagger}\otimes\mathbb{I}_{R}\nonumber\\
 & -a_{k_{1},L}^{\dagger}\otimes a_{k_{2},R}-a_{k_{1},L}^{\dagger}\otimes a_{k_{2},R}^{\dagger}+a_{k_{1},L}^{\dagger}a_{k_{2},L}\otimes\mathbb{I}_{R}+a_{k_{1},L}^{\dagger}a_{k_{2},L}^{\dagger}\otimes\mathbb{I}_{R}\biggr]a_{k',L}^{\dagger}\left|0\right\rangle \otimes\left|0\right\rangle \nonumber\\
 & =\frac{\gamma}{4}\sqrt{\frac{L_{0}}{C_{0}}}\int_{0}^{\pi}\int_{0}^{\pi}\frac{dk_{1}dk_{2}}{(2\pi)^{2}}\frac{\cos\left(\frac{k_{1}}{2}\right)\cos\left(\frac{k_{2}}{2}\right)}{\sqrt{\sin\left(\frac{k_{1}}{2}\right)\sin\left(\frac{k_{2}}{2}\right)}}\left[-\left(2\pi\right)^{2}\left(\delta(k_{2}-k')\delta(k_{1}-k)+\delta(k_{1}-k')\delta(k_{2}-k)\right)\right]\nonumber\\
 & =-\frac{\gamma}{2}\sqrt{\frac{L_{0}}{C_{0}}}\frac{\cos\left(\frac{k}{2}\right)\cos\left(\frac{k'}{2}\right)}{\sqrt{\sin\left(\frac{k}{2}\right)\sin\left(\frac{k'}{2}\right)}}
\end{align}
Hence to first order we get
\begin{align}
\left(G_{k,k'}\right)_{RL} & =-\frac{\gamma}{2}\sqrt{\frac{L_{0}}{C_{0}}}\frac{1}{E-\omega_{k}+i0^{+}}\frac{\cos\left(\frac{k}{2}\right)\cos\left(\frac{k'}{2}\right)}{\sqrt{\sin\left(\frac{k}{2}\right)\sin\left(\frac{k'}{2}\right)}}\frac{1}{E-\omega_{k'}+i0^{+}}\ .
\end{align}
Let us connect the above Green's function to a Green's function in
position space, using the following transformation
\begin{align}
\left|j\right\rangle  & =\int_{0}^{\pi}\frac{dk}{\pi}\cos\left(k\left(j-1/2\right)\right)\left|k\right\rangle
\end{align}
The position space Green's function is
\begin{align}
\left\langle 0,j\right|\frac{1}{E-H+i0^{+}}\left|j',0\right\rangle  & =-\frac{\gamma}{2}\sqrt{\frac{L_{0}}{C_{0}}}\int_{0}^{\pi}\int_{0}^{\pi}\frac{dkdk'}{\pi^{2}}\cos\left(k\left(j-1/2\right)\right)\left(G_{k,k'}\right)_{RL}\cos\left(k'\left(j'-1/2\right)\right)
\end{align}
Hence the transmission coefficient using Eq.~\eqref{eq:Fisher_Lee_formula} is
\begin{align}
t^{L} & =iv\frac{\gamma}{2}\sqrt{\frac{L_{0}}{C_{0}}}\lim_{j\to\infty,j'\to-\infty}\left\langle 0,j\right|\frac{1}{E-H_{0}+i0^{+}}\left|j',0\right\rangle e^{-ik_{E}(j-j')}\nonumber\\
 & =iv\frac{\gamma}{8}\sqrt{\frac{L_{0}}{C_{0}}}\lim_{j\to\infty}\left(\int_{0}^{\pi}\frac{dk}{\pi}\frac{e^{-ik/2}e^{i\left(k-k_{E}\right)j}+e^{ik/2}e^{-i\left(k+k_{E}\right)j}}{E-2\omega_{0}\sin\left(\frac{k}{2}\right)+i0^{+}}\frac{\cos\left(k/2\right)}{\sqrt{\sin\left(k/2\right)}}\right)\nonumber\\
 & *\lim_{j'\to-\infty}\left(\int_{0}^{\pi}\frac{dk'}{\pi}\frac{e^{-ik'/2}e^{i\left(k'+k_{E}\right)j'}+e^{ik'/2}e^{-i\left(k'-k_{E}\right)j'}}{E-2\omega_{0}\sin\left(\frac{k'}{2}\right)+i0^{+}}\frac{\cos\left(k'/2\right)}{\sqrt{\sin\left(k'/2\right)}}\right)
\end{align}
where $k_{E}$ is the wave vector corresponding to the energy at which the signal travels, $E=2\omega_{0}\sin(k_{E}/2)$. Let us
focus on the first integral
\begin{align}
\int_{0}^{\pi}\frac{dk}{\pi}\frac{e^{-ik/2}e^{i\left(k-k_{E}\right)j}+e^{ik/2}e^{-i\left(k+k_{E}\right)j}}{E-2\omega_{0}\sin\left(\frac{k}{2}\right)+i0^{+}}\frac{\cos\left(k/2\right)}{\sqrt{\sin\left(k/2\right)}}\approx e^{-ik_{E}j}\int_{0}^{\pi}\frac{dk}{\pi}\frac{e^{-ik/2}e^{ikj}}{E-2\omega_{0}\sin\left(\frac{k}{2}\right)+i0^{+}}\frac{\cos\left(k/2\right)}{\sqrt{\sin\left(k/2\right)}}
\end{align}
where we have ignored the term $e^{-i(k+k_{E})j}$ as it will become
highly oscillatory in the limit $j\to\infty$. We furthermore consider energies sufficiently low, such that we can also make a linear
approximation for the sine and cosine functions
\begin{align}
e^{-ik_{E}j}\int_{0}^{\pi}\frac{dk}{\pi}\frac{e^{-ik/2}e^{ikj}}{E-2\omega_{0}\sin\left(\frac{k}{2}\right)+i0^{+}}\frac{\cos\left(k/2\right)}{\sqrt{\sin\left(k/2\right)}}\approx\frac{e^{-ik_{E}j}}{\omega_{0}}\int_{0}^{\infty}\frac{dk}{\pi}\frac{e^{ikj}}{k_{E}-k+i0^{+}}\sqrt{\frac{2}{k}}\ .
\end{align}
This is in accordance with taking the continuum limit, i.e., the dimensions of the islands $j$ approaching zero.

To evaluate the remaining integral, we perform a contour integral
in complex $k$-space, where the contour is a quarter circle in the
first quadrant centered at the origin and with radius $R$, hence
\begin{align}
\oint_{C}\frac{dz}{\pi}\frac{e^{izj}}{k_{E}-z+i0^{+}}\sqrt{\frac{2}{z}} & =\int_{0}^{R}\frac{dk}{\pi}\frac{e^{ikj}}{k_{E}-k+i0^{+}}\sqrt{\frac{2}{k}}+\int_{\theta=0}^{\pi/2}\frac{iRe^{i\theta}d\theta}{\pi}\frac{e^{ijR(\cos\theta+i\sin\theta)}}{k_{E}-Re^{i\theta}+i0^{+}}\sqrt{\frac{2}{Re^{i\theta}}}\nonumber\\
 & +\int_{R}^{0}\frac{idk}{\pi}\frac{e^{-kj}}{k_{E}-ik+i0^{+}}\sqrt{\frac{2}{ik}}\ .
\end{align}
The left-hand side of the above can be evaluated using Residue theorem, to
get
\begin{align}
\oint_{C}\frac{dz}{\pi}\frac{e^{izj}}{k_{E}-z+i0^{+}}\sqrt{\frac{2}{z}} & =\frac{2\pi i}{\pi}\lim_{z\to k_{E}+i0^{+}}\left(z-\left(k_{E}+i0^{+}\right)\right)\frac{e^{izj}}{k_{E}+i0^{+}-z}\sqrt{\frac{2}{z}}\nonumber\\
 & =-2ie^{i\left(k_{E}+i0^{+}\right)j}\sqrt{\frac{2}{k_{E}+i0^{+}}}\nonumber\\
 & =-2ie^{ik_{E}j}\sqrt{\frac{2}{k_{E}}}
\end{align}
To evaluate the right-hand side, we take the limit $R\to\infty$, hence the angular
integral goes to zero, which gives us
\begin{align}
\int_{0}^{\infty}\frac{dk}{\pi}\frac{e^{ikj}}{k_{E}-k+i0^{+}}\sqrt{\frac{2}{k}} & =\sqrt{2i}\int_{0}^{\infty}\frac{dk}{\pi}\frac{e^{-kj}}{k_{E}-ik}\sqrt{\frac{1}{k}}-2ie^{ik_{E}j}\sqrt{\frac{2}{k_{E}}}\nonumber\\
 & =\sqrt{2}*\frac{ie^{ijk_{E}}\text{erfc}\left(\sqrt{j}\sqrt{ik_{E}}\right)}{\sqrt{k_{E}}}-2ie^{ik_{E}j}\sqrt{\frac{2}{k_{E}}} \ .
\end{align}
Hence
\begin{align}
\int_{0}^{\pi}\frac{dk}{\pi}\frac{e^{-ik/2}e^{i\left(k-k_{E}\right)j}+e^{ik/2}e^{-i\left(k+k_{E}\right)j}}{E-2\omega_{0}\sin\left(\frac{k}{2}\right)+i0^{+}}\frac{\cos\left(k/2\right)}{\sqrt{\sin\left(k/2\right)}} & \approx e^{-ik_{E}j}\left(ie^{ijk_{E}}\text{erfc}\left(\sqrt{j}\sqrt{ik_{E}}\right)\sqrt{\frac{2}{k_{E}}}-2ie^{ik_{E}j}\sqrt{\frac{2}{k_{E}}}\right)\nonumber\\
 & =i\text{erfc}\left(\sqrt{j}\sqrt{ik_{E}}\right)\sqrt{\frac{2}{k_{E}}}-2i\sqrt{\frac{2}{k_{E}}}\ .
\end{align}
Similarly for the second integral we get
\begin{equation}
\int_{0}^{\pi}\frac{dk'}{\pi}\frac{e^{-ik'/2}e^{i\left(k'+k_{E}\right)j'}+e^{ik'/2}e^{-i\left(k'-k_{E}\right)j'}}{E-2\omega_{0}\sin\left(\frac{k'}{2}\right)+i0^{+}}\frac{\cos\left(k'/2\right)}{\sqrt{\sin\left(k'/2\right)}}\approx i\text{erfc}\left(\sqrt{-j'}\sqrt{ik_{E}}\right)\sqrt{\frac{2}{k_{E}}}-2i\sqrt{\frac{2}{k_{E}}}\ .
\end{equation}
Finally the transmission coefficient is
\begin{align}
t^{L} & =iv\frac{\gamma}{8\omega_{0}^{2}}\sqrt{\frac{L_{0}}{C_{0}}}\lim_{j\to\infty}\left(i\text{erfc}\left(\sqrt{j}\sqrt{ik_{E}}\right)\sqrt{\frac{2}{k_{E}}}-2i\sqrt{\frac{2}{k_{E}}}\right)\lim_{j'\to-\infty}\left(i\text{erfc}\left(\sqrt{-j'}\sqrt{ik_{E}}\right)\sqrt{\frac{2}{k_{E}}}-2i\sqrt{\frac{2}{k_{E}}}\right)\nonumber\\
 & =-i\frac{\gamma L_{0}}{k_{E}}\ ,
\end{align}
where $v=\omega_{0}\cos(k_{E}/2)\approx\omega_{0}$. The error function
parts in both the integrals goes to zero in the respective limits. We thus arrive at Eq.~\eqref{eq:transmission_coeff} in the main text.

\end{widetext}

\bibliography{biblio_AJ_JS_RR}

\begin{thebibliography}{126}
\bibitem [{foo()}]{foot_1}%
  \BibitemOpen
  \href@noop {} {}\bibinfo {note} {Note however, that we expect our conclusions
  qualitatively hold also for other systems providing a single-level charge
  level in between the superconductors, such as a superconducting charge island
  (Cooper pair transistor in the regime of large charging energies, see, e.g.,
  \cite{Cottet_thesis}).}\BibitemShut {Stop}%
\end{thebibliography}

\end{document}